\documentclass[aps,pre,twocolumn,showpacs,floatfix,groupedaddress,nofootinbib]{revtex4}

\usepackage{algorithm}
\usepackage{algpseudocode}
\usepackage{amsmath}
\usepackage{amsthm}
\usepackage{amssymb}
\usepackage{amscd}
\usepackage{bm}
\usepackage{latexsym}
\usepackage{mathtools}
\usepackage{comment}

\usepackage{graphicx}
\usepackage{epsfig}
\usepackage{epstopdf}
\usepackage{svg}
\usepackage[caption=false]{subfig}
\usepackage{placeins}
\usepackage{tabularx}

\usepackage{url}
\usepackage{hyperref}

\renewcommand{\vec}[1]{\mathbf{#1}}

\providecommand{\diff}{\mathrm{d}}

\providecommand{\abs}[1]{\vert#1\vert}
\providecommand{\norm}[1]{\Vert#1\Vert}

\providecommand{\eqref}[1]{(\ref{#1})}

\providecommand{\vanish}[1]{}

\usepackage{lipsum}

\begin{document}
	
\title{Foundations of a Finite Non-Equilibrium Statistical Thermodynamics:\\ Extrinsic Quantities}

\author{O. B. Eri\c{c}ok}
\affiliation{Materials Science and Engineering, University of California, Davis, CA, 95616, USA.}

\author{J. K. Mason}
\email{jkmason@ucdavis.edu}
\affiliation{Materials Science and Engineering, University of California, Davis, CA, 95616, USA.}

\begin{abstract}

Statistical thermodynamics is valuable as a conceptual structure that shapes our thinking about equilibrium thermodynamic states. A cloud of unresolved questions surrounding the foundations of the theory could lead an impartial observer to conclude that statistical thermodynamics is in a state of crisis though. Indeed, the discussion about the microscopic origins of irreversibility has continued in the scientific community for more than a hundred years. This paper considers these questions while beginning to develop a statistical thermodynamics for finite non-equilibrium systems. Definitions are proposed for all of the extrinsic variables of the fundamental thermodynamic relation that are consistent with existing results in the equilibrium thermodynamic limit. The probability density function on the phase space is interpreted as a subjective uncertainty about the microstate, and the Gibbs entropy formula is modified to allow for entropy creation without introducing additional physics or modifying the phase space dynamics. Resolutions are proposed to the mixing paradox, Gibbs' paradox, Loschmidt's paradox, and Maxwell's demon thought experiment. Finally, the extrinsic variables of the fundamental thermodynamic relation are evaluated as functions of time and space for a diffusing ideal gas, and the initial and final values are shown to coincide with the expected equilibrium values when interpreted in a classical context.

\end{abstract}

\maketitle

\section{Introduction}
\label{sec:introduction}

While thermodynamics is nominally concerned with heat and energy flows, the only states about which anything meaningful can be said about are those infinitesimally close to equilibrium, i.e., those for which heat and energy flows vanish. Moreover, being based on empirical observations, thermodynamics has little to say about the underlying causes for observed phenomena. That is instead the subject of statistical thermodynamics, which has the purpose of establishing connections between the microscopic state of a system (i.e., the particle positions and momenta) and equilibrium thermodynamic quantities. Although statistical thermodynamics can rightfully claim a number of significant successes, the theory says little about non-equilibrium systems by design.

That is instead the concern of non-equilibrium statistical thermodynamics. This subject has received considerable attention (perhaps for the reason that the most physically significant systems are often those not in equilibrium) but is not nearly as well-established as equilibrium statistical thermodynamics. This can be partly attributed to the difficulty of adequately defining and measuring thermodynamic quantities for a non-equilibrium system; without established experimental points of reference, it is difficult to conclude whether any given proposal should be discarded as being discordant with reality. Part of the difficulty is also that there are ontological questions that have not been satisfactorily resolved. For example, can a thermodynamic entropy be defined for a finite isolated system, one of precisely the type that is frequently simulated by molecular dynamics?

There are at least three practical reasons to search for answers to such a question. First, students and practitioners who are asked to accept the conclusions of statistical thermodynamics without finding satisfactory answers to such questions could easily lose confidence in a subject that is, and should be regarded as, one of the cornerstones of modern physics. Second, the inconsistencies that emerge when pondering such questions could erroneously suggest that our instinctive sense of reality is not relevant to this subject; without this, our ability to reason about and develop non-equilibrium statistical thermodynamics is severely limited. Third, the possible applications of molecular dynamics to the physical sciences would be considerably expanded if thermodynamic quantities like the free energy could be reliably extracted from molecular dynamics simulations.

The purpose of this article is to establish part of the foundations for a non-equilibrium statistical thermodynamics that is consistent with existing results in the equilibrium thermodynamic limit, and that introduces a minimum of arbitrary constants and additional physics. Section \ref{sec:limitations} motivates this undertaking by considering several limitations of the existing foundations, including situations where the ergodic hypothesis does not hold, the dependence of thermodynamic ensembles on initial conditions, and the increase of entropy during irreversible processes. Section \ref{sec:fundamental} proposes definitions applicable to finite non-equilibrium systems of all of the extrinsic quantities in the fundamental thermodynamic relation. These definitions satisfy the same properties as the classical quantities in the equilibrium thermodynamic limit. Section \ref{sec:ideal_gas} evaluates these quantities as functions of time and space for the expansion of an ideal gas in an isolated box, and verifies that our definitions converge to the expected values when interpreted in a classical context.

\section{Limitations of Statistical Thermodynamics}
\label{sec:limitations}

The following limitations in the foundations of statistical thermodynamics are not intended to be exhaustive, but rather to provide motivation for the developments in subsequent sections.

\subsection{Thermodynamic Ensembles}
\label{subsec:ensembles}

The utility of a thermodynamic ensemble derives from the ergodic hypothesis \cite{boltzmann1871einige,ehrenfest1990conceptual}, or the effective assumption that the time average of a quantity over a generic trajectory through the phase space can be replaced by the instantaneous average of the quantity over an ensemble of systems prepared in the same macrostate. Practically speaking, the ability to replace a complicated calculation of the evolution of a microstate with an average over the phase space, weighted by an appropriately constructed probability distribution, is the only reason that statistical thermodynamic calculations are at all possible. That said, a variety of systems do not behave ergodically \cite{hahn1950spin,kolmogorov1954conservation,gallavotti2007fermi,dumas2014kam}, and for these the above procedure leads to obvious inconsistencies.

\begin{figure}
	\center
	\includegraphics[width=1.0\linewidth]{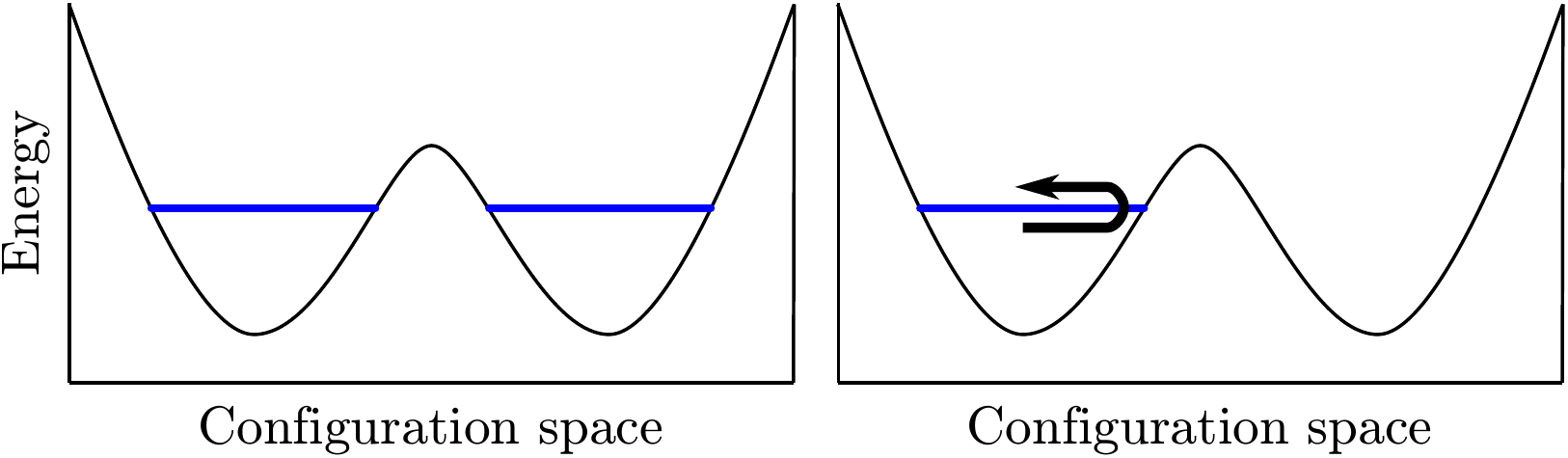} 
	\caption{\label{fig:disconnected}A schematic showing the confinement of a microcanonical system to a small part of the configuration space. The microcanonical ensemble (left) stipulates that all microstates of a given total energy (blue, with the potential energy in black) be equally probable. For a physical system (right), the trajectory is confined to the path-connected region containing the initial condition.}
\end{figure}

As an initial example, consider a microcanonical system containing two potential energy wells, as shown in Fig.\ \ref{fig:disconnected}. The microcanonical ensemble stipulates that, following the principle of equal \emph{a priori} probabilities \cite{mcquarrie1976,tolman1979principles}, the time average of an observable for a microcanonical system with total energy in blue is equivalent to the average over the allowed states in both potential energy wells. That said, the allowed states are separated by a disallowed region, any trajectory is necessarily confined to the potential energy well specified by the initial conditions (i.e., the system is not metrically transitive \cite{birkhoff1931proof,neumann1932proof}), and the ergodicity hypothesis does not hold. The two-dimensional square lattice Ising model with zero external field below the critical temperature is a nontrivial system that exhibits such behavior \cite{onsager1944crystal,yang1952spontaneous}, and shows that this situation is not purely hypothetical. This leads to the natural conclusion that the time-averaged behavior of a generic system can depend on the initial conditions, even in the long-time limit.

One could object that it is not really possible to isolate a physical system from its surroundings, and that ergodicity is restored by the thermal fluctuations in a canonical system eventually allowing the trajectory to pass over the potential energy barrier. The characteristic time required for such a fluctuation depends exponentially on the height of the potential energy barrier though, meaning that the mean residence time in a potential energy well could easily be longer than the lifetime of the universe. More generally, whether the time average of an observable should be performed over one potential energy well or two depends on whether the observation time is shorter or longer than the mean residence time. That is, the time-averaged behavior of a generic system apparently can depend not only on the initial conditions, but also on the period of observation.

Further evidence that an ensemble average is not generally equivalent to the finite time-averaged behavior of a system is offered by the phenomenon of supercooling. Consider a system of supercooled water prepared without any heterogenous crystal nucleation sites. For any specified observation time, let the temperature be such that the system has equal probabilities of homogeneously nucleating a crystal and of remaining a supercooled liquid. The ensemble average would include equal numbers of liquid and crystal states, whereas the time-average behavior of a single system would depend stochastically on whether crystallization occurs in that system. The natural objection to this thought experiment is that thermodynamic ensembles are only relevant to systems in thermodynamic equilibrium, and for that reason are not applicable to a supercooled liquid. That is precisely the point; thermodynamic ensembles have well-established utility but cannot be used in every situation, and care should be taken that the underlying assumptions are not violated.

For the final example, suppose that a phase is defined by a probability distribution on the phase space that is a smooth function of the thermodynamic variables, and that phase transitions occur when this probability distribution changes discontinuously. Consider the proposed existence of an ideal glass phase as a resolution to Kauzmann's paradox \cite{kauzmann1948nature,gibbs1958nature}. Cooling an ensemble of glass forming systems through the glass transition temperature implies that the probability distribution associated with a proposed ideal glass phase should be distributed over all metabasins in the potential energy landscape. This is inconsistent with the observation that the instantaneous configuration of a single glassy system identifies one or a few metabasins to which the system is confined for reasonable observation times. That is, the actual probability distribution of microstates for a single glassy system is inconsistent with the one that would be characteristic of an ideal glass phase, though confusion about this point could occur as a result of inappropriately applying thermodynamic ensembles to a non-equilibrium system.

\subsection{Probability Distributions}
\label{subsec:probability}

This discussion raises questions about the interpretation of a probability distribution of microstates. Given a system in thermodynamic equilibrium, the usual interpretation is that if the practitioner were to repeatedly measure the microstate at intervals longer than the decorrelation time of the system, then the microstates would be independently and identically distributed according to this probability distribution. This requires that the probability distribution be stationary, i.e., the partial derivative with respect to time vanishes everywhere. Along with the constraints imposed by the Liouville equation and the requirement that the weighted average of thermodynamic variables over the phase space be equal (to within a small uncertainty) to the experimentally-measured values, this is conjectured to make the probability distribution well-defined \cite{cornfeld2012ergodic}.

Actually calculating a probability distribution in this way would be exceedingly difficult though. Instead, the usual approach is to construct an ensemble of systems with different initial conditions subject to the requirement that, perhaps after an initial transient, every system in the ensemble realizes a probability distribution of microstates the same as that of the single system described above \cite{boltzmann1871einige,maxwell1879on}. The probability distribution can then be empirically constructed by examining the microstates of the ensemble at any single instant in time after the initial transient. This has the advantage of removing the dependence on the decorrelation time, but still requires that the probability distribution be time invariant and consistent with the macrostate.

The interpretation of a probability distribution of microstates for a non-equilibrium system is less clear. Since the properties of a non-equilibrium system (and therefore the probability distribution of microstates) are not generally constant in time, a practitioner cannot reliably construct an empirical probability distribution from a time series of microstates. As for the construction of a thermodynamic ensemble, requiring the observation time to be longer than the decorrelation time is no longer sufficient for ensemble averages to be independent of the choice of initial microstates (as shown by the thought experiment below), and a thermodynamic ensemble cannot be constructed without making this choice explicit.

Consider a thermodynamic ensemble corresponding to any equilibrium state. The time evolution of the systems in the ensemble maps to an ensemble of trajectories in the phase space. If the practitioner makes an instantaneous change to a thermodynamic variable defining the macrostate, the systems in the ensemble would no longer be in a state of equilibrium. A time-dependent empirical probability distribution for the non-equilibrium state could then be constructed from the ensemble of trajectories, and would be equivalent to the result of applying the Liouville equation to the initial probability distribution. The difficulty with this construction is that, as pointed out in Sec.\ \ref{subsec:ensembles}, thermodynamic ensembles are not generally applicable to non-equilibrium systems. If the initial equilibrium system was a glass former in the liquid phase, instantaneously quenching the system through the glass transition temperature would cause the systems in the corresponding ensemble to break into distinct populations confined to separate metabasins in the potential energy landscape. The actual system would follow a trajectory into one of these metabasins, with the result that the probability distribution of microstates for the actual system would be inconsistent with the prediction of the ensemble.

This conflict is resolved by recognizing that the probability distribution of microstates actually represents the practitioner's subjective uncertainty about the system's microstate \cite{jaynes1968prior,jaynes1965gibbs}. The practitioner does not precisely know the initial microstates of the system nor of the surroundings, and therefore does not know the system's subsequent trajectory. Without this information, it is quite reasonable to represent the initial uncertainty by means of a probability distribution and to propagate that uncertainty to subsequent times with the Liouville equation. That the uncertainty is not an objective property of a thermodynamic system should not affect the apparent properties of generic equilibrium states, since for these there is assumed to be a limiting probability distribution that is independent of the initial conditions. This certainly does not apply to all thermodynamic states though, the clearest examples being ones close to a phase transition where the time averages of distinct trajectories can diverge.

A possible objection to the conclusion of the preceding paragraph is that this makes the thermodynamic entropy a subjective quantity, though the consequences of this are less severe than could be supposed. Consider that the part of the internal energy that cannot feasibly be extracted as work should be expected to increase with the practitioner's uncertainty about the microstate. This allows irreversible processes, e.g., the expansion of an ideal gas into an empty volume, to be interpreted as those where the practitioner's uncertainty about the system's microstate spontaneously increases. Conversely, an entity with perfect knowledge of a system's microstate would conclude that the subjective entropy was zero and be able to extract all of the internal energy as work, allowing Maxwell's demon to avoid violating any thermodynamic laws.

Jaynes' discussion of the mixing paradox offers further evidence that the entropy should be considered as a subjective quantity \cite{jaynes1992gibbs}. Consider a thermodynamic system consisting of two chambers separated by a removable barrier, the first containing gas $A$ and the second containing gas $B$. If the practitioner is able to distinguish gases $A$ and $B$, then the entropy of the system should increase with the removal of the barrier and the mixing of the gases. Conversely, if the practitioner cannot distinguish gases $A$ and $B$, then the removal of the barrier is reversible and should not change the entropy of the system. The mixing paradox arises by considering what happens when the physical properties of gases $A$ and $B$ are continuously varied from a condition where they are distinguishable to one where they are indistinguishable given the capabilities of the practitioner. While the physical properties of the system change continuously, the entropy change on removing the barrier is apparently discontinuous at the point where the gasses first become indistinguishable. This would be concerning if the entropy were an objective property of the system, but if the entropy is subjective this merely reflects that the practitioner's knowledge of the possible microstates changes discontinuously.

The subjectivity of the probability distribution allows a significant reinterpretation of the discussion of supercooling in Sec.\ \ref{subsec:ensembles}. After placing a liquid in a freezer, the practitioner does not know whether the system has crystallized or not until the freezer door is opened. The probability distribution describing the practitioner's knowledge of the system state at this point includes equal numbers of liquid and crystal states; this accurately reflects that it is not possible to reliably predict the work that could be extracted from the system, or more generally the system's responses to external stimuli, until more information is acquired. Opening the door provides this information and collapses the probability distribution into one containing only liquid or only crystal states, once again allowing useful predictions to be made. While this phenomenon superficially resembles the wavefunction collapse of quantum mechanics, it is emphasized that the act of observation does not in any way affect the objective trajectories of the molecules in the liquid, merely the practitioner's knowledge of and ability to harness the system. Thermodynamics is then not so much about describing physical systems as it is about describing the ways in which physical systems can be manipulated.

\subsection{Liouville's Theorem}
\label{subsec:liouville}

Given an initial probability distribution on the phase space, the Liouville equation describes the time evolution of this probability distribution on the basis of Hamiltonian mechanics \cite{liouville1838note,gibbs1902elementary}; conceptually, the flow of probability density resembles that of an incompressible fluid. Liouville's theorem states that the convective derivative of the probability distribution is zero, or that any set of points beginning within a specified volume can always be enclosed in a distorted region of identical volume at any subsequent time. Suppose that the thermodynamic entropy is given by Gibbs's entropy formula for a continuous probability distribution
\begin{equation}
	S_\mathrm{G} = -k_\mathrm{B} \int_{\Gamma} \rho \ln \! \frac{\rho}{\mu} \diff \Omega
	\label{eq:gibbs_entropy}
\end{equation}
where $\Gamma$ is the phase space, $\diff \Omega$ is the infinitesimal volume element, $\rho$ is the probability distribution, and $\mu$ is a constant that makes the argument of the logarithm unitless. Liouville's theorem then implies that Gibbs's entropy is constant for any time evolution given by the Liouville equation, even for initial conditions that describe non-equilibrium systems. For this reason, it is unclear how the second law of thermodynamics can be derived from statistical thermodynamics without the introduction of additional physics.

Only two of the resolutions proposed in the literature are considered here. The first is known as coarse-graining, and involves partitioning the phase space into regions of a characteristic volume $V_\mathrm{CG}$ \cite{gibbs1902elementary,tolman1979principles,penrose2005foundations}. This allows the continuous probability distribution $\rho$ to be converted into a discrete probability distribution $q_i$ indexed by the partition elements. Gibbs's entropy formula for a continuous probability distribution is then converted into with one for a discrete probability distribution (up to an additive constant):
\begin{equation*}
	S_\mathrm{CG} = -k_\mathrm{B} \sum_i q_i \ln q_i.
\end{equation*}
This approach has several advantages. With regard to the second law, the increasing distortion of a volume element in the phase space with time results in a continuous probability distribution being distributed over many elements of the partition, allowing the coarse-grained entropy to increase despite Liouville's theorem. More precisely, the introduction of $V_\mathrm{CG}$ defines a length scale below which the contraction of a continuous probability distribution cannot be resolved, and after an initial transient the rate of expansion of the continuous probability distribution effectively depends only on the positive Lyapunov exponents \cite{eckmann1985ergodic,pesin1991characteristic}. With regard to the third law, the concentration of a continuous probability distribution at a single point as the temperature approaches absolute zero would cause the entropy in Eq.\ \ref{eq:gibbs_entropy} to diverge to negative infinity, but the coarse-grained entropy approaches the conventional value of zero for any nonzero $V_\mathrm{CG}$.

\begin{figure}
	\center
	\includegraphics[width=0.6\linewidth]{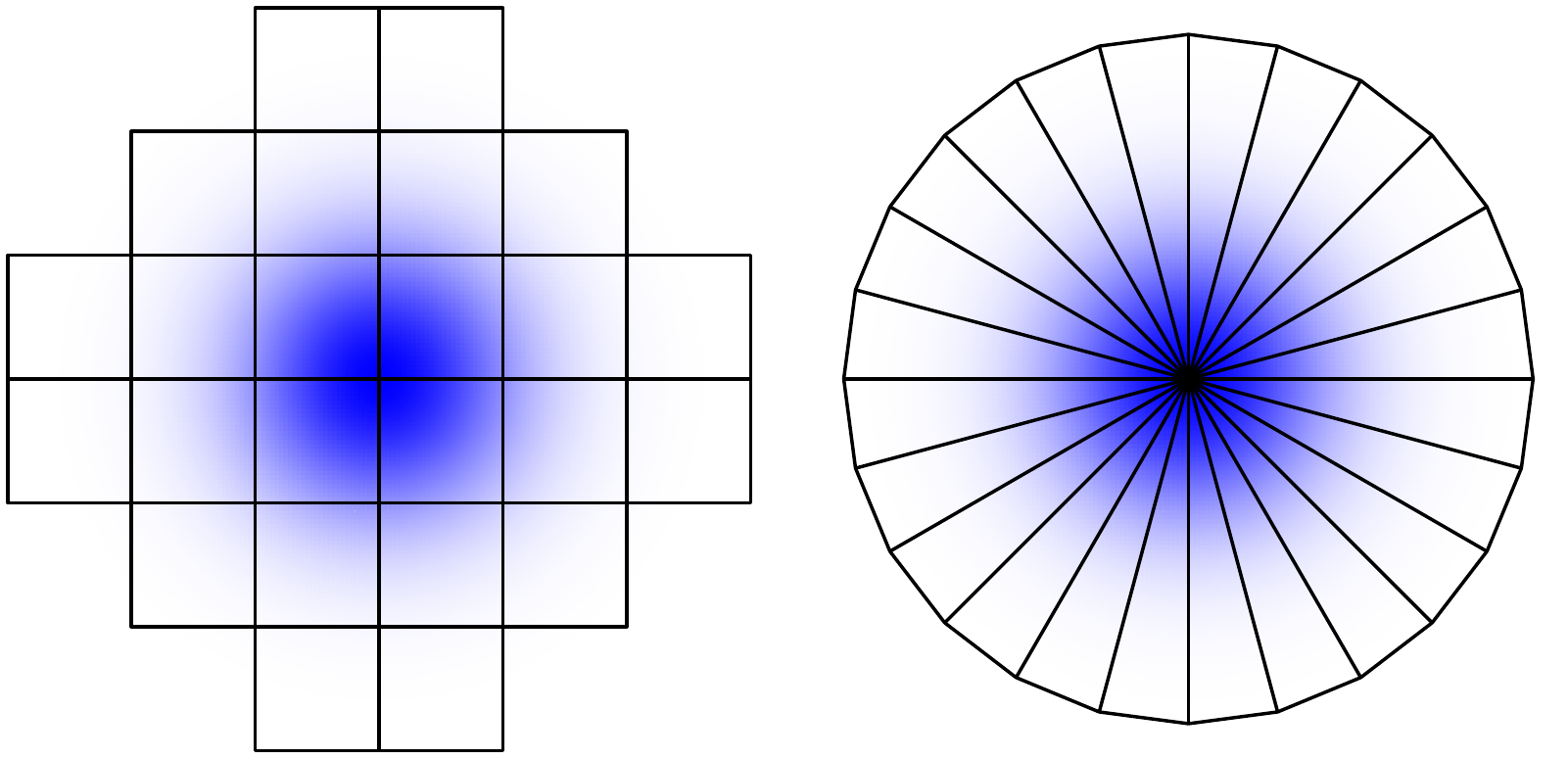} 
	\caption{\label{fig:figure2}Representations of two partitions of the phase space containing a fixed number of elements, with the probability distribution in blue. The right partition maximizes the coarse-grained entropy.}
\end{figure}

Everything comes with a price though, and for the coarse-graining approach that is introducing $V_\mathrm{CG}$ as a fundamental physical constant that behaves as an offset to the entropy. Consider that the coarse-grained entropy for a uniform continuous probability distribution on some volume $V_\Gamma$ of the phase space is
\begin{align*}
	S_\mathrm{CG} &= -k_\mathrm{B} \sum_i \frac{V_\mathrm{CG}}{V_\Gamma} \ln \! \frac{V_\mathrm{CG}}{V_\Gamma} \\
	&= k_\mathrm{B} \ln \! \frac{V_\Gamma}{V_\mathrm{CG}}.
\end{align*}
Since taking the $V_\mathrm{CG} \rightarrow 0$ limit causes the coarse-grained entropy to diverge, every thermodynamic system should have a nonzero $V_\mathrm{CG}$. Moreover, for thermodynamic comparisons to be possible, this value should be the same for all thermodynamic systems. The definition of a fundamental physical constant is not something to be undertaken lightly though, and the frequent identification of $V_\mathrm{CG}$ with the appropriate power of Planck's constant is inconsistent with the correspondence principle and the development of a classical statistical thermodynamics. Finally, there remains the problem of the construction of the partition of phase space. Consider that it is possible to construct, for any continuous probability distribution and $V_\mathrm{CG}$, a partition for which the coarse-grained entropy is equal to that of the uniform continuous probability distribution over the volume $V_\Gamma$; this idea is illustrated in Fig.\ \ref{fig:figure2}. This would clearly be inconsistent with thermodynamic measurements, but in the absence of a physically-justified construction such a choice is not forbidden.

\begin{figure}
	\center
	\includegraphics[width=0.5\linewidth]{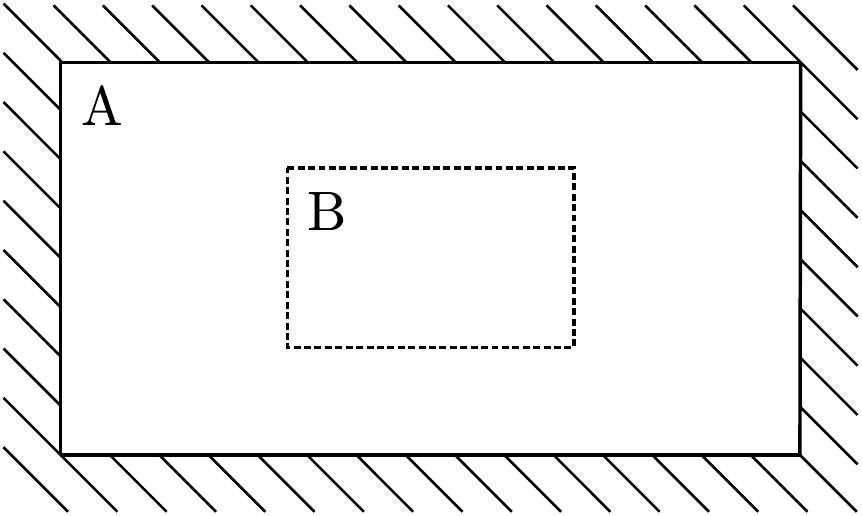} 
	\caption{\label{fig:figure3}The thermodynamic system $A$ is assumed to be isolated, and the closed subsystem $B$ is separated from the surroundings $A - B$ by a heat-conductive barrier.}
\end{figure}

The second proposed resolution instead modifies the Liouville equation by introducing a diffusive term, with the result that an initial probability distribution spreads over time and Gibbs's entropy in Eq.\ \ref{eq:gibbs_entropy} increases \cite{blatt1959alternative,lindblad2001non,ridderbos2002coarse}. The usual justification offered is that it is not actually possible to construct an isolated system, and the diffusive term reflects the effectively stochastic influence of the surroundings. The consequences of this argument can be seen by considering the isolated system $A$ in Fig.\ \ref{fig:figure3}; it should always be possible to construct such a system since, e.g., the universe is isolated by definition. Let the initial probability distribution of microstates of $A$ be such that the entropy $S_\mathrm{G}(A)$ is equal to the minimum possible for any thermodynamic system. Since $A$ does not interact with any surroundings, the probability distribution of microstates of $A$ obeys the standard Liouville equation and $S_\mathrm{G}(A)$ remains constant by Liouville's theorem. If a subsystem $B \subset A$ is closed but not isolated, the probability distribution of microstates of $B$ obeys the modified Liouville equation, and the entropy $S_\mathrm{G}(B)$ gradually increases to a limiting value. Experimental thermodynamics requires that the entropy be additive in the sense that $S_\mathrm{G}(A) = S_\mathrm{G}(B) + S_\mathrm{G}(A - B)$ though, and if $S_\mathrm{G}(B) > S_\mathrm{G}(A)$, then $S_\mathrm{G}(A - B) < S_\mathrm{G}(A)$ must be less than the absolute minimum of entropy. That is, the proposed modification of the Liouville equation apparently violates the third law of thermodynamics.

Proponents of coarse-graining could suggest that a regular tessellation of the phase space by hypercubes be used as a canonical partition, and perhaps average over all possible translations of the tessellation. Proponents of the modified Liouville equation could object that the universe is not isolated, or that the situation in the paragraph above could not occur in practice. There are many other proposed answers to the question of entropy generation in the literature as well. The purpose of this section is not to rigorously evaluate these proposals, but merely to point out that they generally seem to require additional development before they could be entirely satisfactory.

\section{Fundamental Thermodynamic Relation}
\label{sec:fundamental}

The development of classical thermodynamics begins with the fundamental thermodynamic relation for open systems
\begin{equation}
	\diff U = T \diff S + V \sum_{ij} \sigma_{ij} \diff \epsilon_{ij} + \sum_i \mu_i \diff N_i.
	\label{eq:fundamental_classical}
\end{equation}
The purpose of this section is to propose suitable definitions of the extrinsic quantities in Eq.\ \ref{eq:fundamental_classical} that can be applied to finite non-equilibrium systems and that reduce to the classical results in the equilibrium limit. This necessarily includes a definition of the entropy that is consistent with the second and third laws of thermodynamics. Our intention is to handle the intrinsic quantities in a separate publication.

It is useful to define precisely what properties are expected of an extensive quantity, particularly since the literature is surprisingly inconsistent on this point \cite{touchette2002quantity,mannaerts2014extensive}. Extensive quantities have been variously defined as those that are equal to the sum of the corresponding values for the constituent subsystems (additive), as those that are proportional to a system's mass (proportional), or as those that are are both additive and proportional. These definitions are not all equivalent, and are not necessarily sensible for non-equilibrium systems. Consider that the volume of a system composed of two subsystems with different densities but the same number of atoms (e.g., a liquid and a solid) is additive but not proportional. The gravitational potential energy of a system increases superlinearly in the number of atoms, implying that the internal energy is not proportional either.

Classical definitions of extensivity apparently make several assumptions that are usually left implicit. First, the extent of the interactions between any two subsystems should be at most proportional to the shared surface area. Second, passing to the thermodynamic limit causes each subsystem's volume to increase faster than its surface area, making interactions between subsystems negligible in the limit. A classically extensive quantity is here defined to be one that is additive in such conditions. Notice that the second assumption is not applicable to the finite systems considered here.

Two types of additivity are defined for finite systems that are equivalent to the classical definition in the conditions described above. A quantity is said to be \emph{strongly additive} if its value for a union of arbitrary subsystems is equal to the sum of its values for those subsystems. Strong additivity is stronger than classical additivity since it does not require either of the assumptions above, and is often a property of quantities that are defined on a per-atom basis. A quantity is said to be \emph{weakly additive} if its value for a union of isolated subsystems (defined in Sec.\ \ref{subsec:subsystems}) is equal to the sum of its values for those subsystems. Notice that the assumptions above effectively isolate each subsystem, making strong and weak additivity equivalent in classical conditions.

The quantities defined in the sections below are said to be strongly or weakly extensive depending on whether they are strongly or weakly additive; this will be found to depend sensitively on the way the subsystems' interactions are handled. Finally, it is worthwhile to briefly discuss whether one is preferable to the other. Certainly they are both consistent with macroscale experimental thermodynamics, and so one does not seem to be preferable to the other on that account. It is true that strong extensivity is mathematically convenient, but it is not at all clear that it should be possible to objectively distribute, e.g., the energy of an interaction over the various subsystems participating in that interaction. Observe that since the energy in question arises from interacting subsystems, the part of the energy assigned to a particular subsystem cannot be a property of that subsystem alone. Indeed, the appealing idea that a subsystem's internal energy should in some way reside within that subsystem was an essential part of the intuitive but long-discredited caloric theory \cite{morris1972lavoisier}.

\subsection{Boundary Conditions}
\label{subsec:subsystems}

Given that one of our motivations is to evaluate thermodynamic quantities during molecular dynamics simulations, it is useful to consider the ways in which such a system could be defined. Since molecular dynamics simulations are constructed to conserve internal energy, volume, and particle number unless specifically specified otherwise, the simplest simulation would be of an \emph{isolated} system for which all of these quantities are constant and Eq.\ \ref{eq:fundamental_classical} is the natural governing equation.

If instead the system is \emph{closed} and can exchange only thermal energy with a reservoir, then that is precisely what should be simulated. This would entail a molecular dynamics simulation of a large isolated system $A$ partitioned into a system of interest $B$ and a surrounding reservoir $A - B$. $B$ would need to be separated from $A - B$ by a solid heat-conductive barrier composed of atoms with a high binding energy. Notice that the usual practice of neglecting boundary effects in the thermodynamic limit is not valid for finite systems, and that thermodynamic quantities should include the effects of the system's particles interacting with those of the barrier. The use of a thermostat is inadvisable since it is not yet clear that the definition of temperature for an equilibrium system is relevant to a non-equilibrium one.

For an \emph{open} system that can exchange thermal energy and particles with a reservoir, the procedure would be similar but with the barrier removed. This raises the question of what precisely constitutes the system of interest $B$ and the surrounding reservoir $A - B$ though. Since the particles in $B$ are not fixed, $B$ would presumably be defined so as to include the particles residing within a particular region of space at any instant in time. Notice that defining the thermodynamic properties of $B$ using the probability distribution of microstates of $B$ leads to difficulty since the dimension of the phase space would change discontinuously with the number of particles in $B$. For this reason, the thermodynamic properties of $B$ should always be defined using the probability distribution of microstates of $A$.

The only change necessary to allow a closed system to exchange volume with a reservoir would be to make the heat-conductive barrier deformable. It is unclear whether a barrier could be made deformable while disallowing the transfer of thermal energy for a system that can exchange only volume with a reservoir though. A simulation could be performed by subjecting the system to a barostat, but it is not yet clear that the definition of pressure for an equilibrium system is relevant to a non-equilibrium one.

There is not an obvious way to define what constitutes an open system that is able to exchange volume with a reservoir, since neither the particles in $B$ nor the region assigned to $B$ are fixed. Other possible boundary conditions that involve the exchange of particles but not thermal energy with a reservoir are regarded as unphysical.

\subsection{Internal Energy}
\label{subsec:internal_energy}

The internal energy is the first of the extrinsic quantities appearing in Eq.\ \ref{eq:fundamental_classical}, and is defined as the sum of the potential and kinetic energies
\begin{equation*}
	U(\vec{x}, \vec{p}) = U_{x}(\vec{x}) + U_{p}(\vec{p})
\end{equation*}
where $\vec{x}$ and $\vec{p}$ are vectors of all the position and momentum variables describing the microstate of the system. Since energy is not an absolute quantity, the zeros of the potential and kinetic energies need to be defined. For atomic systems with short-range interactions, the potential energy is usually measured with respect to a microstate where the atoms are infinitely separated and in the absence of external fields, and the kinetic energy is measured with respect to a center of momentum frame of the system. While $U_{x}(\vec{x})$ is usually a complicated function of the position variables, $U_{p}(\vec{p})$ is simply the sum of the kinetic energies of the constituent atoms.

The internal energy above is a function on the phase space that assigns a value to a particular microstate, but the expectation value $\langle U \rangle = \langle U_{x} \rangle + \langle U_{p} \rangle$ of the practitioner depends on the probability distribution of microstates $\rho(\vec{x}, \vec{p})$. Since the potential and kinetic energies respectively depend only on position and momentum variables, their expectation values can be calculated as
\begin{align*}
	\langle U_{x} \rangle &= \int_{\Gamma_x} U_{x}(\vec{x}) \rho_x(\vec{x}) \diff \vec{x} \\
	\langle U_{p} \rangle &= \int_{\Gamma_p} U_{p}(\vec{p}) \rho_p(\vec{p}) \diff \vec{p}
\end{align*}
where $\rho_x = \int_{\Gamma_p} \rho \, \diff \vec{p}$ and $\rho_p = \int_{\Gamma_x} \rho \, \diff \vec{x}$ are the marginal probability distributions on the configuration and momentum spaces $\Gamma_x$ and $\Gamma_p$. The change in internal energy is the more physically relevant quantity though. Suppose that the probability distribution $\rho$ is changed into the probability distribution $\rho + \delta \rho$, where $\delta \rho(\vec{x}, \vec{p})$ is arbitrarily small with respect to the $L^{1}$ norm and is required to satisfy $\int_{\Gamma} \delta \rho \, \diff \Omega = 0$ for the probability distribution to remain normalized. The resulting change $\diff \langle U \rangle = \diff \langle U_{x} \rangle + \diff \langle U_{p} \rangle$ in the expectation value of the internal energy is the sum of the changes in the expectation values of the potential and kinetic energies, or
\begin{align*}
	\diff \langle U_{x} \rangle &= \int_{\Gamma_x} U_{x}(\vec{x}) \delta \rho_x(\vec{x}) \diff \vec{x} \\
	\diff \langle U_{p} \rangle &= \int_{\Gamma_p} U_{p}(\vec{p}) \delta \rho_p(\vec{p}) \diff \vec{p}
\end{align*}
where $\delta \rho_x = \int_{\Gamma_p} \delta \rho \, \diff \vec{p}$ and $\delta \rho_p = \int_{\Gamma_x} \delta \rho \, \diff \vec{x}$ are defined analogously to $\rho_x$ and $\rho_p$. This is in principle enough to define the change in internal energy of an isolated system.

Physical systems are rarely isolated though. Even noble gas atoms in a rigid container interact by dispersion forces with the atoms in the container walls, and there should be a procedure to partition the potential energy of such interactions into two parts, one pertaining to the system and the other to the surroundings. More generally, let a system $A$ be partitioned into $n$ subsystems $A^{i}$ where $i \in [1, n]$. A point in the phase space $\Gamma$ defines the microstate of $A$ and therefore the microstates of all $A^{i}$. This implies that there should exist functions $U^{i}_{x}(\vec{x})$ and $U^{i}_{p}(\vec{p})$ defined on $\Gamma$ that indicate the potential and kinetic energies of $A^{i}$ and that have expectation values defined analogously to $\langle U_{x} \rangle$ and $\langle U_{p} \rangle$. Moreover, the extensivity of the internal energy requires that these functions satisfy $U_{x}(\vec{x}) = \sum_i U^{i}_{x}(\vec{x})$ and $U_{p}(\vec{p}) = \sum_i U^{i}_{p}(\vec{p})$.

A strongly extensive $U^{i}_{p}(\vec{p})$ can be defined by assigning to $A^{i}$ the kinetic energies of all the atoms in $A^{i}$, or
\begin{equation}
	U_{p}^{i}(\vec{p}) = \sum_{j \in \mathcal{L}^{i}} \norm{\vec{p}_{j}}^2 / (2 m_j)
	\label{eq:kinetic_energy_part}
\end{equation}
where $\mathcal{L}^{i}$ is the set of labels of the atoms in $A^{i}$ and $\vec{p}_{j}$ is the momentum of the $j$th atom. Defining $U^{i}_{x}(\vec{x})$ is more difficult though. Let $N$ be the number of atoms in $A$, and $\mathcal{S}$ be the set of all subsets of labels of atoms in $A$ except for the empty set. Formally, the potential energy of $A$ can be represented as a sum of many-body interactions $U_{x}(\vec{x}) = \sum_{s \in \mathcal{S}} \phi_{s}(\vec{x})$ where $\phi_{s}(\vec{x})$ is the interaction among the atoms with labels in $s$ \cite{hankins1970water}. A reasonable definition for $U_{x}^{i}(\vec{x})$ is then
\begin{equation}
	U_{x}^{i}(\vec{x}) = \sum_{s \in \mathcal{S}} \frac{\abs{\mathcal{L}^{i} \cap s}}{\abs{s}} \phi_{s}(\vec{x})
	\label{eq:potential_energy_part}
\end{equation}
where the contribution of an interaction to the potential energy of a subsystem is proportional to the number of atoms of the subsystem that participate in the interaction. It is straightforward to show that this definition is strongly extensive.

The difficulty with this approach is that it is not clear whether the potential energy can always be decomposed into a sum of many-body interactions in practice; electronic structure calculations indicate that the presence of one atom can affect the electron distributions of surrounding atoms in complicated ways. That said, it is always possible to measure the potential energy change of removing an atom (i.e., displacing the atom to infinity). The question is therefore whether such information is sufficient to construct $U^{i}_{x}(\vec{x})$ as in the definition above.

Observe that $\mathcal{S}$ not only indexes the many-body interactions, but also the sets of atoms that could be removed from the system. Let $\chi_{r}(\vec{x})$ be the potential energy change of $A$ when removing the atoms with labels in $r \in \mathcal{S}$. By construction, $\chi_{r}(\vec{x})$ is the sum of all many-body interactions that involve any atoms with labels in $r$, and is written as $\chi_{r}(\vec{x}) = \sum_{s \in \mathcal{S}} a_{rs} \phi_{s}(\vec{x})$ where $a_{rs} \in \{-1,0\}$. The matrix with elements $a_{rs}$ is invertible as shown by the following argument. The atom removal event that includes all of the atoms of $A$ gives the potential energy of $A$. Adding the potential energy of $A$ to that of events that remove all but one atom gives the one-body interactions. Adding the potential energy of $A$ to that of events that remove all but two atoms and subtracting the appropriate one-body interactions gives the two-body interactions. Repeating this process with increasing numbers of remaining atoms shows that each atom removal event introduces precisely one multi-body interaction that is not already known. The rows of the matrix with elements $a_{rs}$ are therefore linearly independent, and since the matrix is square, it is invertible. If $b_{sr}$ are the elements of the inverse matrix, the multi-body interactions can be found by $\phi_{s}(\vec{x}) = \sum_{r \in \mathcal{S}} b_{sr} \chi_{r}(\vec{x})$. Substituting this into Eq.\ \ref{eq:potential_energy_part} gives
\begin{equation}
	U_{x}^{i}(\vec{x}) = \sum_{s \in \mathcal{S}} \frac{\abs{\mathcal{L}^{i} \cap s}}{\abs{s}} \sum_{r \in \mathcal{S}} b_{sr} \chi_{r}(\vec{x})
	\label{eq:potential_energy_part2}
\end{equation}
An example of this procedure is given in App.\ \ref{sec:partitioning_potential_energy}.

Since the number of multi-body interactions increases as $2^{N}$, the definition in Eq.\ \ref{eq:potential_energy_part2} could be difficult to use in practice when the number of atoms is not sufficiently small. As an alternative, the potential energy of a multi-body interaction could be distributed equally over the subsystems that participate in the interaction. Let $S_{n}$ be the set of all permutations of $n$ elements, $\sigma \in S_{n}$ be a permutation that indicates an ordering of the subsystems, and consider the process that begins with the system $A$ and successively removes all atoms in $A^{\sigma(j)}$ for $j \in [1, n]$. Let $\psi^{i}_{\sigma}(\vec{x})$ be the potential energy change when the atoms in $A^{i}$ are removed as part of this process. Then
\begin{equation}
    U^{i}_{x}(\vec{x}) = \frac{1}{n!} \sum_{\sigma \in S_{n}} \psi^{i}_{\sigma}(\vec{x})
    \label{eq:potential_energy_part3}
\end{equation}
is a weakly extensive definition for the potential energy of $A^{i}$. A proof that this effectively distributes many-body interactions equally over the participating subsystems is given in App.\ \ref{sec:extrinsic_formula}. The advantage of this definition is that if an isolated system $A$ is partitioned into a subsystem $A^{1}$ of interest and the surroundings $A^{2}$, the equation for $U^{1}_{x}(\vec{x})$ has only two terms and Eq.\ \ref{eq:potential_energy_part3} could be evaluated in practice.

One property of these definitions is that if every atom of $A$ is assigned to a separate subsystem, Eqs.\ \ref{eq:potential_energy_part2} and \ref{eq:potential_energy_part3} give identical and well-defined potential energies for individual atoms. The alternatives in the literature instead require the careful definition and justification of volumes associated with individual atoms over which an energy density is integrated \cite{bader1994atoms,yu2011energy}.

\subsection{Entropy}
\label{subsec:entropy}

If the definitions for the extrinsic quantities in the fundamental thermodynamic relation are to reduce to the classical ones in the equilibrium limit, then the entropy is likely to be defined by an equation resembling Gibbs's formula in Eq.\ \ref{eq:gibbs_entropy}. Section \ref{subsec:liouville} describes why this appears to not allow entropy generation during irreversible processes as required by the second law of thermodynamics though. This suggests that the entropy formula be modified, though it is useful to initially consider the symmetries that the entropy is expected to satisfy. Certainly the entropy, like all other thermodynamic quantities, should be invariant to the choice of coordinate system. The entropy in Eq.\ \ref{eq:gibbs_entropy} not only has this property, but is actually invariant to all canonical transformations (ones that preserve the form of Hamilton's equations of motion) \cite{landau1976mechanics,goldstein1980classical}. These include transformations that, e.g., exchange coordinates with momenta, though since other thermodynamic variables are not expected to be invariant to such transformations, it is possible that the entropy in Eq.\ \ref{eq:gibbs_entropy} is more symmetric than is necessary.

Following this line of thought, it is curious that position and momentum variables enter equivalently into Eq.\ \ref{eq:gibbs_entropy}, simply as variables over which to integrate. They are certainly not equivalent in the context of Lagrangian mechanics, nor in that of quantum mechanics \cite{landau2013quantum,shankar2012principles}. Indeed, quantum mechanics even seems to suggest that the marginal probability distributions $\rho_{x}$ and $\rho_{p}$ are more fundamental objects than the joint probability distribution $\rho$. Suppose then that the entropy is defined as a sum of configurational and vibrational entropies
\begin{equation}
	\langle S \rangle = \langle S_{x} \rangle + \langle S_{p} \rangle.
	\label{eq:modified_entropy}
\end{equation}
Applying Gibbs's formula to the marginal probability distributions $\rho_{x}$ and $\rho_{p}$ allows these to be explicitly defined as
\begin{align*}
	\langle S_{x} \rangle &= -k_\mathrm{B} \int_{\Gamma_x} \rho_{x}(\vec{x}) \ln \! \frac{\rho_{x}(\vec{x})}{\mu_{x}} \diff \vec{x} \\
	\langle S_{p} \rangle &= -k_\mathrm{B} \int_{\Gamma_p} \rho_{p}(\vec{p}) \ln \! \frac{\rho_{p}(\vec{p})}{\mu_{p}} \diff \vec{p}.
\end{align*}
The quantum entropic uncertainty principle \cite{1975beckner,1975bialynicki} states that, for suitable values of $\mu_{x}$ and $\mu_{p}$, the entropy in Eq.\ \ref{eq:modified_entropy} can be made nonnegative for any $\rho_{x}$ and $\rho_{p}$ that are admissible on the basis of quantum mechanics; while the intention is to develop a classical theory, it is satisfying that the entropy remains sensible even in the low-temperature limit where quantum effects are likely to be significant.

The entropies in Eqs.\ \ref{eq:gibbs_entropy} and \ref{eq:modified_entropy} are equivalent whenever $\vec{x}$ and $\vec{p}$ are independent random variables, or $\rho = \rho_x \rho_p$, since
\begin{align}
	-\frac{\langle S \rangle}{k_\mathrm{B}} &= \int_{\Gamma_x} \rho_{x}(\vec{x}) \ln \! \frac{\rho_{x}(\vec{x})}{\mu_{x}} \diff \vec{x} + \int_{\Gamma_p} \rho_{p}(\vec{p}) \ln \! \frac{\rho_{p}(\vec{p})}{\mu_{p}} \diff \vec{p} \nonumber \\
	&= \int_\Gamma \rho(\vec{x}, \vec{p}) \ln \! \frac{\rho_{x}(\vec{x})}{\mu_{x}} \diff \Omega + \int_\Gamma \rho(\vec{x}, \vec{p}) \ln \! \frac{\rho_{p}(\vec{p})}{\mu_{p}} \diff \Omega \nonumber \\
	&= \int_\Gamma \rho(\vec{x}, \vec{p}) \ln \! \frac{\rho_{x}(\vec{x}) \rho_{p}(\vec{p})}{\mu_{x} \mu_{p}} \diff \Omega.
	\label{eq:modified_entropy_explicit}
\end{align}
Significantly, the Gibbs measure for the canonical distribution has this property, and can be written as the product of marginal measure on the configuration space and a Maxwell--Boltzmann distribution on the momentum space. The definition in Eq.\ \ref{eq:modified_entropy} is therefore equivalent to that in Eq.\ \ref{eq:gibbs_entropy} for the canonical ensemble, and by the principle of the equivalence of ensembles \cite{ellis2000large,touchette2015equivalence}, is consistent with effectively all prior thermodynamic results in the equilibrium thermodynamic limit. This includes the Sackur--Tetrode equation \cite{sackur1911anwendung,tetrode1912chemische}, one of the few explicit predictions for the entropy of a thermodynamic system. That said, and despite the entropy often being separated into configurational and vibrational parts in the literature, the two definitions of entropy are not equivalent in general; their difference is proportional to the mutual information of $\vec{x}$ and $\vec{p}$ (here the mutual information of two disjoint sets of random variables is defined analogously to that of two random variables \cite{cover2012elements}).

With regard to the second law, the entropy in Eq.\ \ref{eq:modified_entropy} generically increases with time for a non-equilibrium isolated system essentially as a result of projective geometry. Since the Lyapunov exponents of a Hamiltonian system occur in pairs $(\lambda, -\lambda)$ \cite{dettmann1996proof,abraham2008foundations}, a generic initial probability distribution expands exponentially in some directions and contracts exponentially in others. Though Liouville's theorem indicates that the overall volume of this region is constant in time, the projected volumes on the configuration and momentum spaces are not, as can be seen by considering the projected shadow of an immiscible fluid stirred into water. More precisely, the evolution of $\rho$ as prescribed by the Liouville equation generically causes the mutual information of $\vec{x}$ and $\vec{p}$ and therefore the entropy in Eq.\ \ref{eq:modified_entropy} to increase with time up to some limiting value when the system is said to have reached equilibrium.

The notion of a generic initial probability distribution should be examined further. If the entropy increase required by the second law is provided by an increase in mutual information, then a decrease of entropy with time as in Loschmidt's paradox is not strictly forbidden \cite{loschmidt1876uber,boltzmann1877bemerkungen}. Consider any initial probability distribution $\rho_{0}$ for which the entropy increases in time. After some period of time $\tau$ has passed and $\rho_{0}$ has evolved to $\rho_{1}$, transform $\rho_{1}$ to $\rho'_{1}$ by mapping each point in the phase space to the corresponding point with the particle velocities reversed. The evolution of $\rho'_{1}$ would then proceed back to $\rho'_{0}$ with decreasing entropy. That is, for each initial probability distribution $\rho_{0}$ for which entropy increases over $\tau$, there is a corresponding initial probability distribution $\rho'_{1}$ for which entropy decreases by the same amount over $\tau$. Apparently, a generic initial probability distribution does not refer to most probability measures on the phase space in the usual sense.

Recall the conclusion of Sec.\ \ref{subsec:probability} that the probability distribution reflects the practitioner's uncertainty about the system's microstate; perhaps a generic probability distribution instead refers to those that a practitioner could feasibly specify as an initial condition. Certainly probability distributions with little mutual information (e.g., where $\vec{x}$ and $\vec{p}$ are independent random variables) are simpler to write down and to reason about, but the more fundamental difficulty relates to the possible sources of information about correlations of position and momentum variables. It seems unlikely that the practitioner would be able to either prepare an initial condition with highly correlated position and momentum variables or measure such correlations in an evolving system. Again consider stirring an immiscible fluid into water. Certainly the most likely initial condition from a practical standpoint is one where the immiscible fluid is a roughly spherical droplet, i.e., where the marginal mass distributions along orthogonal directions are nearly uncorrelated. Moreover, once stirring begins the distribution of mass quickly becomes so complicated that the details are practically unavailable, and the only quantity that can reasonably be measured is the mass density. This suggests that the resolution to Loschmit's paradox is that the available initial conditions do not include all possible measures on the phase space due to the limitations of human capabilities and human knowledge.

With regard to symmetries, the entropy defined in Eq.\ \ref{eq:modified_entropy} is invariant to coordinate transformations but not to all canonical transformations. This would not be true if the reference quantities $\mu_{x}$ and $\mu_{p}$ were not included with the appropriate units of inverse volume though, since then the absolute value of the Jacobian determinant would appear inside the logarithms during a coordinate transformation. The presence of $\mu_{x}$ and $\mu_{p}$ is therefore essential, and their values can be determined by requiring that the entropy vanish at the absolute zero of temperature. For simplicity, consider the entropy of a perfect crystal in the Einstein model \cite{einstein1907plancksche,mandl2013statistical}. Let the crystal contain $N$ atoms of mass $m$ and have a potential energy of $U_{x}(\vec{x}) = \frac{1}{2} m \omega^2 \sum_i \abs{\vec{x}_i - \vec{y}_i}^2$, where $\vec{x}_i$ and $\vec{y}_i$ are the absolute position and minimum potential energy position of the $i$th atom. The marginal probability distributions in the ground state can then be shown to be \cite{landau2013quantum,shankar2012principles}
\begin{align*}
	\rho_{x}(\vec{x}) &= \bigg( \frac{m \omega}{\pi \hbar} \bigg)^{3 N / 2} \exp \! \bigg( -\frac{m \omega}{\hbar} \sum_j \abs{\vec{x}_j - \vec{y}_j}^2 \bigg) \\
	\rho_{p}(\vec{p}) &= \bigg( \frac{1}{\pi \hbar m \omega} \bigg)^{3 N / 2} \exp \! \bigg( -\frac{1}{\hbar m \omega} \sum_j \abs{\vec{p}_j}^2 \bigg).
\end{align*}
The third law of thermodynamics is interpreted as requiring that the corresponding configurational and vibrational entropies vanish independently. Further assuming that $\mu_{x}$ and $\mu_{p}$ are constant functions on their respective spaces gives
\begin{align*}
	\mu_{x} &= \bigg( \frac{2 m \omega}{h e} \bigg)^{3 N / 2} \\
	\mu_{p} &= \bigg( \frac{2}{h m \omega e} \bigg)^{3 N / 2}
\end{align*}
where $e$ is Euler's number. While these depend on the choice of reference crystal through $m$ and $\omega$, there is not an obvious physical reason why the entropy of all reference crystals should be the same at the absolute zero of temperature. Certainly one does not expect, e.g., the heat capacities of all reference crystals to be the same in the low-temperature limit. The appearance of Plank's constant could be argued as being inconsistent with a classical theory, but since $\mu_{x}$ and $\mu_{p}$ are defined by systems at the absolute zero of temperature where quantum effects are significant, this is not regarded as a serious objection. Finally, the dependence on $N$ is necessary for $\mu_{x}$ and $\mu_{p}$ to have the appropriate units of inverse volume in the configuration and momentum spaces.

The derivation of the equation for the entropy change $\diff \langle S \rangle = \diff \langle S_{x} \rangle + \diff \langle S_{p} \rangle$ is slightly more involved than the one for the internal energy since the entropy depends nonlinearly on the probability distribution $\rho$. Consider the configurational entropy of a system with probability distribution $\rho + \delta \rho$:
\begin{align*}
	-\frac{\langle S_{x} \rangle'}{k_\mathrm{B}} &= \int_{\Gamma_x} (\rho_{x} + \delta \rho_{x}) \ln \! \frac{\rho_{x} + \delta \rho_{x}}{\mu_{x}} \diff \vec{x} \\
	&= \int_{\Gamma_x} \bigg[ \rho_{x} \ln \! \frac{\rho_{x}}{\mu_{x}} + \delta \rho_{x} \ln \! \frac{\rho_{x}}{\mu_{x}} + \delta \rho_x + \frac{(\delta \rho_{x})^2}{\rho_{x}} \bigg] \diff \vec{x} \\
	&= -\frac{\langle S_{x} \rangle}{k_\mathrm{B}} + \int_{\Gamma_x} \delta \rho_{x} \ln \! \frac{\rho_{x}}{\mu_{x}} \diff \vec{x}.
\end{align*}
The second line uses a first-order Taylor series expansion of the logarithm about $\rho_{x} / \mu_{x}$, and the third line only retains terms to first order in $\delta \rho_{x}$. This allows the changes in the configurational and vibrational entropies to be written as
\begin{align*}
	\diff \langle S_{x} \rangle &= -k_\mathrm{B} \int_{\Gamma_x} \delta \rho_{x}(\vec{x}) \ln \! \frac{\rho_{x}(\vec{x})}{\mu_{x}} \diff \vec{x} \\
	\diff \langle S_{p} \rangle &= -k_\mathrm{B} \int_{\Gamma_p} \delta \rho_{p}(\vec{p}) \ln \! \frac{\rho_{p}(\vec{p})}{\mu_{p}} \diff \vec{p}.
\end{align*}
Curiously, $\diff \langle S_{x} \rangle$ and $\diff \langle S_{p} \rangle$ are independent of $\mu_{x}$ and $\mu_{p}$ despite their appearance in the equations above. Observe that, e.g., multiplying $\mu_{x}$ by an arbitrary constant $a$ is equivalent to subtracting $\ln a \int_{\Gamma_x} \delta \rho_{x} \diff \vec{x}$ from the first integral, but this vanishes since $\int_{\Gamma_x} \delta \rho_{x} \diff \vec{x} = 0$. This indicates that while $\mu_{x}$, $\mu_{p}$, and the absolute entropy might depend on the choice of reference crystal, the validity of the fundamental thermodynamic relation does not.

There is still the issue of the extensivity of the entropy and the frequent division of the partition function by a factor of $N!$ to account for the indistinguishability of identical particles. This practice is apparently motivated by a thought experiment proposed by Gibbs \cite{gibbs1902elementary}. Consider two chambers of the same volume $V$ separated by a removable barrier, each containing $N$ atoms of an ideal monatomic gas at the same temperature. The $\rho_{x}$ of a single chamber is a constant function over a volume $V^{N}$, with the corresponding configurational entropy
\begin{equation*}
	\langle S_{x} \rangle = k_\mathrm{B} \ln (V^{N} \mu^{(N)}_{x})
\end{equation*}
where the superscript on $\mu_{x}$ indicates the number of atoms for which the constant is defined. Removing the barrier results in a single system of volume $2 V$ with $2 N$ atoms for which $\rho_{x}$ is a constant function over a volume $(2V)^{2N}$. The configurational entropy after removing the barrier is therefore
\begin{equation*}
	\langle S_{x} \rangle' = 2 k_\mathrm{B} \ln (V^{N} \mu^{(N)}_{x}) + 2 k_\mathrm{B} N \ln 2
\end{equation*}
where $\mu_{x}$ was squared to account for the change in the dimension of the configuration space. The change in configurational entropy $\Delta \langle S_{x} \rangle = \langle S_{x} \rangle' - 2 \langle S_{x} \rangle$ is then
\begin{equation*}
	\Delta \langle S_{x} \rangle = 2 k_\mathrm{B} N \ln 2
\end{equation*}
which is inconsistent with the second law; since the chambers were initially in equilibrium, the removal of the barrier should be reversible and the entropy change should be zero.

This paradox is usually resolved by postulating that the $N$ gas atoms in a chamber are identical and indistinguishable, reducing the volume of distinct states in the configuration space to $V^{N} / N!$ instead of $V^{N}$ \cite{leinaas1977theory,huang2009introduction}. Repeating the steps above with this modification gives the expected $\Delta \langle S_{x} \rangle \approx 0$ (to a precision limited by Stirling's approximation). This treatment is unsatisfactory for at least three reasons. First, the Liouville equation does not indicate what happens when the dimension of the phase space changes. More precisely, removing the barrier transforms the phase space from a disjoint union of those of the two chambers into a product space of the same, a troublingly discontinuous change given that the physical system evolves continuously. Second, if the physical properties of the gas atoms in one chamber could be varied continuously, the equation for the entropy would change discontinuously at the point when the gases are first distinguishable. This seems less preferable to assigning the discontinuity to the practitioner's uncertainty about the microstate, as was done with the mixing paradox in Sec.\ \ref{subsec:probability}. Third, the modern justification for the indistinguishability of identical atoms relies on quantum mechanical principles, and a classical statistical thermodynamics would ideally depend on quantum mechanics as little as possible.

The change in the dimension of the phase space can be addressed by instead considering the two chambers as parts of a single thermodynamic system at all points in time. If, e.g., the first chamber is known to hold the first $N$ atoms and the second chamber the second $N$ atoms, then $\rho_{x}$ is the product of two independent probability distributions for the atoms in each of the two chambers. This makes $\rho_{x}$ a constant function over a volume $V^{2N}$, and the initial configurational entropy
\begin{equation*}
	\langle S_{x} \rangle = k_\mathrm{B} \ln(V^{2N} \mu^{(2N)}_{x}).
\end{equation*}
Removing the barrier gives the same $\rho_{x}$ and $\langle S_{x} \rangle'$ as above (up to the change in the definition of $\mu_{x}$), with the result that the change in configurational entropy $\Delta \langle S_{x} \rangle = \langle S_{x} \rangle' - \langle S_{x} \rangle$ is
\begin{equation*}
	\Delta \langle S_{x} \rangle = 2 k_\mathrm{B} N \ln 2.
\end{equation*}
An increase in the configurational entropy is expected though, since removing the barrier irreversibly increases the practitioner's uncertainty about the microstate; before removing the barrier the first $N$ atoms were known to be in the first chamber and the second $N$ atoms in the second chamber, but afterwards each atoms has equal probability of being in either chamber. 

While justifiable, this is inconsistent with the expectation that a practitioner would not observe any measurable change in the macrostate on removing and reinserting the barrier. The source of the inconsistency is not the equation for the entropy though, but rather that the initial $\rho_{x}$ does not accurately reflect the practitioner's uncertainty about the microstate. By hypothesis the practitioner knows that the system begins with $N$ atoms in the first chamber and $N$ atoms in the second, but any permutation of the $2 N$ atoms would result in an experimentally indistinguishable system in all respects. The initial $\rho_{x}$ should therefore be a constant function over $(2 N)! / (N!)^2$ copies of a volume $V^{2N}$, one for each binning of the atoms. The corresponding initial configurational entropy is
\begin{align*}
	\langle S_{x} \rangle &= k_\mathrm{B} \ln(V^{2N}\mu^{(2N)}_{x}) + k_\mathrm{B} \ln \! \frac{(2N)!}{(N!)^2} \\
	&\approx k_\mathrm{B} \ln( V^{2N}\mu^{(2N)}_{x}) + 2 k_\mathrm{B} N \ln 2
\end{align*}
where the second line uses Stirling's approximation, with the consequence that $\Delta \langle S_{x} \rangle \approx 0$. That is, the extensivity of the entropy does not require that identical atoms actually be indistinguishable (a postulate not obviously compatible with a classical theory), but merely that they be effectively indistinguishable as a consequence of the limited knowledge of the practitioner.

Following the discussion in Sec.\ \ref{subsec:subsystems}, it would be useful to be able to distribute the entropy over several subsystems while maintaining extensivity. As with the internal energy in Sec.\ \ref{subsec:internal_energy}, suppose that a system $A$ is partitioned into $n$ subsystems $A^{i}$ where $i \in [1, n]$. Let $\mathcal{T}$ be the set of all assignments of the $N$ atoms in $A$ to the $n$ subsystems, and associate with the assignment $t \in \mathcal{T}$ the part of the configuration space in which all the atoms are in the indicated subsystems. Given assignment $t$, let $N^{i}_{t}$ be the number of atoms in $A^{i}$, and $\vec{x}^{i}_{t}$ and $\vec{p}^{i}_{t}$ be the sets of position and momentum variables for the atoms in $A^{i}$. Starting with Eq.\ \ref{eq:modified_entropy_explicit}, the contribution of the region associated with assignment $t$ to the overall entropy is
\begin{align*}
	-\frac{\langle S \rangle_{t}}{k_\mathrm{B}} &= 
	\int\limits_{\Gamma_{p}} \int\limits_{A^{n}} \ldots \int\limits_{A^{1}}
	\rho(\vec{x}, \vec{p}) \ln \! \frac{\rho_{x}(\vec{x}) \rho_{p}(\vec{p})}{\mu_{x} \mu_{p}} 
	\diff \vec{x}^{1}_{t} \ldots \diff \vec{x}^{n}_{t} \diff \vec{p} \\
	& = \int\limits_{A^{n}} \ldots \int\limits_{A^{1}}
	\rho_{x}(\vec{x}) \ln \! \frac{\rho_{x}(\vec{x})}{\mu_{x}}
	\diff \vec{x}^{1}_{t} \ldots \diff \vec{x}^{n}_{t} \\
	& \quad + \int\limits_{\Gamma_{p}} \Bigg[ \int\limits_{A^{n}} \ldots \int\limits_{A^{1}}
	\rho(\vec{x}, \vec{p})
	\diff \vec{x}^{1}_{t} \ldots \diff \vec{x}^{n}_{t} \Bigg] \ln \! \frac{\rho_{p}(\vec{p})}{\mu_{p}} \diff \vec{p}
\end{align*}
where the domain of integration for each of the $N^{i}_{t}$ atoms in $A^{i}$ is $A^{i}$. The additive decomposition on the second line suggests the definitions
\begin{align}
	\langle S_{x} \rangle_{t} &= -k_\mathrm{B} \int\limits_{A^{n}} \ldots \int\limits_{A^{1}} \rho_{x}(\vec{x}) \ln \! \frac{\rho_{x}(\vec{x})}{\mu_{x}} \diff \vec{x}^{1}_{t} \ldots \diff \vec{x}^{n}_{t} \label{eq:entropy_assignment} \\
	\langle S_{p} \rangle_{t} &= -k_\mathrm{B} \int\limits_{\Gamma_{p}} \Bigg[ \int\limits_{A^{n}} \ldots \int\limits_{A^{1}} \rho(\vec{x}, \vec{p})\diff \vec{x}^{1}_{t} \ldots \diff \vec{x}^{n}_{t} \Bigg] \ln \! \frac{\rho_{p}(\vec{p})}{\mu_{p}} \diff \vec{p} \nonumber 
\end{align}
for the contributions of the assignment $t$ to the configurational and vibrational entropies of $A$. 

The procedure to distribute the quantities $\langle S_{x} \rangle_{t}$ and $\langle S_{p} \rangle_{t}$ over the subsystems $A^{i}$ is motivated by the chain rule for conditional entropies \cite{cover2012elements}. Let $S_{n}$ be the set of all permutations of $n$ elements, $\sigma \in S_{n}$ be a permutation that indicates an ordering of the subsystems, and $q$ be the number of elements before $i$ in $\sigma$. Let
\begin{align*}
    \alpha^{i}_{t}(\vec{x} | \sigma) &= \int_{\Gamma_{x}} \ldots \int_{\Gamma_{x}} \rho_{x}(\vec{x}) \diff \vec{x}^{\sigma(1)}_{t} \ldots \diff \vec{x}^{\sigma(q)}_{t} \\
    \beta^{i}_{t}(\vec{p} | \sigma) &= \int_{\Gamma_{p}} \ldots \int_{\Gamma_{p}} \rho_{p}(\vec{p}) \diff \vec{p}^{\sigma(1)}_{t} \ldots \diff \vec{p}^{\sigma(q)}_{t}
\end{align*}
be the marginal probability distributions for atoms in subsystems with labels in the final $n - q$ elements of $\sigma$. The quantities
\begin{align*}
    \zeta^{i}_{t}(\vec{x}) &= \frac{1}{n!} \sum_{\sigma \in S_{n}} \ln \Bigg[ \frac{\alpha^{i}_{t}(\vec{x} | \sigma)}{\mu^{(N^{i}_{t})}_{x} \int_{\Gamma_{x}} \alpha^{i}_{t}(\vec{x} | \sigma) \diff \vec{x}^{i}_{t}} \Bigg] \\
    \eta^{i}_{t}(\vec{p}) &= \frac{1}{n!} \sum_{\sigma \in S_{n}} \ln \Bigg[ \frac{\beta^{i}_{t}(\vec{p} | \sigma)}{\mu^{(N^{i}_{t})}_{p} \int_{\Gamma_{p}} \beta^{i}_{t}(\vec{p} | \sigma) \diff \vec{p}^{i}_{t}} \Bigg]
\end{align*}
effectively average the conditional entropies of the atoms in $A^{i}$ over all possible orderings of the subsystems. This gives
\begin{align}
	\langle S^{i}_{x} \rangle_{t} &= -k_\mathrm{B} \int\limits_{A^{n}} \ldots \int\limits_{A^{1}}
	\rho_{x}(\vec{x}) \zeta^{i}_{t}(\vec{x})
	\diff \vec{x}^{1}_{t} \ldots \diff \vec{x}^{n}_{t} \nonumber \\
	\langle S^{i}_{p} \rangle_{t} &= -k_\mathrm{B} \int\limits_{\Gamma_{p}} \Bigg[ \int\limits_{A^{n}} \ldots \int\limits_{A^{1}}
	\rho(\vec{x}, \vec{p})
	\diff \vec{x}^{1}_{t} \ldots \diff \vec{x}^{n}_{t} \Bigg] \eta^{i}_{t}(\vec{p}) \diff \vec{p}
	\label{eq:entropy_partition_parts}
\end{align}
for the configurational and vibrational entropies of $A^{i}$ for assignment $t$. Observe that summing $\langle S^{i}_{x} \rangle_{t}$ and $\langle S^{i}_{p} \rangle_{t}$ over all subsystems gives $\langle S_{x} \rangle_{t}$ and $\langle S_{p} \rangle_{t}$ as defined in Eq.\ \ref{eq:entropy_assignment}. Finally,
\begin{equation}
	\langle S^{i} \rangle = \sum_{t \in \mathcal{T}} (\langle S^{i}_{x} \rangle_{t} + \langle S^{i}_{p} \rangle_{t}).
	\label{eq:entropy_partition}
\end{equation}
is defined as the entropy of subsystem $A^{i}$. This definition is weakly extensive; since the microstates of isolated subsystems are necessarily independently distributed, $\zeta^{i}_{t}(\vec{x})$ and $\eta^{i}_{t}(\vec{p})$ reduce to functions of $\vec{x}^{i}_{t}$ and $\vec{p}^{i}_{t}$ alone, and $\langle S^{i}_{x} \rangle_{t}$ and $\langle S^{i}_{p} \rangle_{t}$ are weakly additive.

The procedure above is for subsystems that are open and defined by particular regions of space. If the subsystems are instead defined by sets of atoms, the only change is that there is a single assignment $t \in \mathcal{T}$. If the subsystems are able to exchange volume, the domain of each subsystem is extended to that of $A$. These changes significantly simplify the calculations for subsystems that are isolated or closed.

\subsection{Volume}
\label{subsec:volume}

If the system is not specified by a region of space but rather by a set of atoms, then the volume likely should be calculated as the expectation value of the sum of the atomic volumes. Only the overall system volume is considered here, the intention being to discuss the strains along with the stresses in a separate publication.

Let $\mathcal{L}$ the set of labels of atoms in the system, $R_{i}(\vec{x})$ be the Voronoi cell of atom $i$, and $|R_{i}(\vec{x})|$ be the volume of this cell. The proposed definition for the volume is then
\begin{equation*}
    \langle V \rangle = \int_{\Gamma_{x}} \rho_{x}(\vec{x}) \sum_{i \in \mathcal{L}} |R_{i}(\vec{x})| \diff \vec{x},
\end{equation*}
though this is likely sensible only when the distribution of the atoms is relatively homogeneous. It is straightforward to show that this definition is strongly extensive.

\subsection{Particle Number}
\label{subsec:number}

The final extensive variables in Eq.\ \ref{eq:fundamental_classical} are the particle numbers $N_{i}$. As discussed in Sec.\ \ref{subsec:subsystems}, the system should be open and specified by a region of space for the calculation of this quantity to be necessary. Let $N_{i}(\vec{x})$ be the number of atoms of species $i$ within the specified region for a given microstate. Then
\begin{equation}
	\langle N_{i} \rangle = \int_{\Gamma_{x}} \rho_{x}(\vec{x}) N_{i}(\vec{x}) \diff \vec{x}
	\label{eq:expected_number_per_partition}
\end{equation}
is the proposed definition for the number of atoms of species $i$ in the system, and significantly is a continuous function of the probability distribution on the configuration space. As with the other definitions that can be written as sums of per-atom quantities, this definition is strongly extensive.

\section{Application to an Ideal Gas}
\label{sec:ideal_gas}

This section applies the definitions developed in Sec.\ \ref{sec:fundamental} to the expansion of an ideal gas in an isolated box. This system that has been studied in the context of non-equilibrium statistical mechanics before \cite{hobson1968exact,kasperkovitz1988approach,espanol1991initial,evans1995long}, though the present approach differs in several respects. For example, Hobson and Loomis \cite{hobson1968exact} considered a system of $N$ ideal gas particles in an isolated box, solved Liouville's equation for the evolution of the probability distribution, and used this to calculate various macroscopic observables like the center of mass, total energy, and total momentum. Then they discarded the previously-derived probability distribution, constructed a different one using the macroscopic observables and Jaynes' principle of maximum uncertainty \cite{jaynes1957information}, and used that to calculate various thermodynamic quantities including the entropy. While they observed that the entropy increased to the expected equilibrium value, this should not be confused for a signature of irreversibility. Instead, this was a result of willfully ignoring previously-known information about the probability distribution. This approach, vocally advocated by Jaynes \cite{jaynes1965gibbs}, has been widely criticized in the literature \cite{lavis1985work,earman1986problem,uffink2011subjective}.

That said, we do use their solution of Liouville's equation for a system of $N$ ideal gas particles initially confined by a removable barrier to one side of a $d$-dimensional isolated box. The system is initially assumed to be in thermodynamic equilibrium with the joint probability distribution $\rho(\vec{x}, \vec{p}, 0) = \rho_{x}(\vec{x}) \rho_{p}(\vec{p})$ where $\rho_{x}(\vec{x})$ is a uniform distribution and $\rho_{p}(\vec{p})$ is a Maxwell--Boltzmann distribution. Removing the barrier at time $t = 0$ allows the particles to diffuse into the rest of the box, with the evolution of the time-dependent probability distribution $\rho(\vec{x}, \vec{p}, t)$ found by the procedure in App.\ \ref{sec:ideal_gas_in_a_box}. Since the nature of an ideal gas allows the solution to be written as a product of the solutions for a single particle in a one-dimensional box, it is the single-particle solution that is analyzed in detail below.

\begin{figure*}
	\center
	\includegraphics[width=0.8\linewidth]{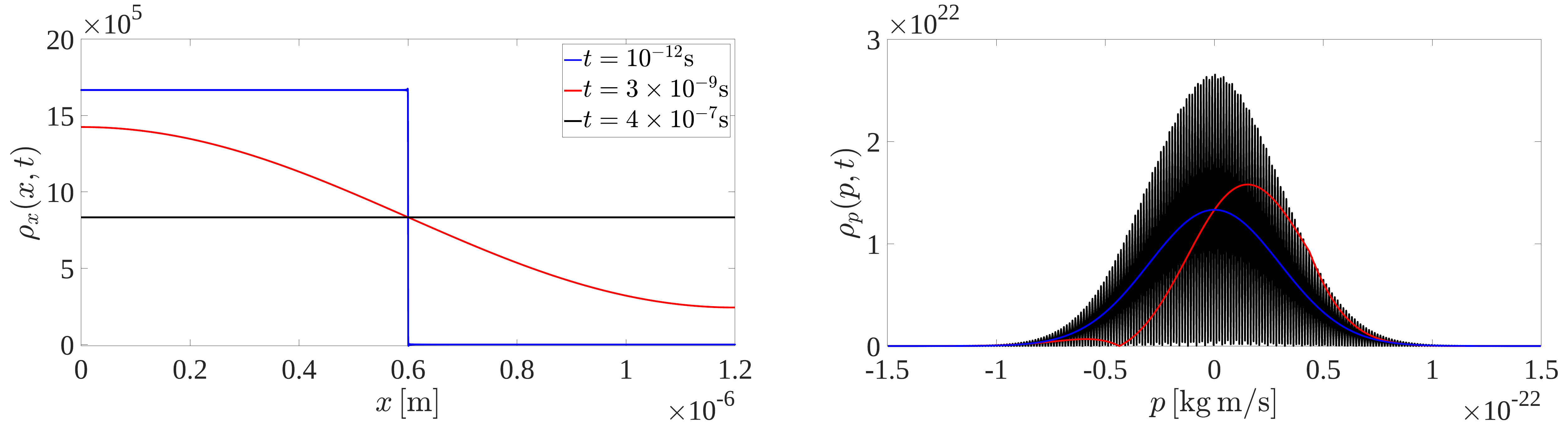}
	\caption{\label{fig:figure4}The evolution of the marginalized probability distributions $\rho_{x}(x,t)$ (left) and $\rho_{p}(p,t)$ (right) for a single xenon atom that is initially confined to the left half of a one-dimensional box of length $l = 1.2 \times 10^{-6} \, \mathrm{m}$ at $T = 298 \, \mathrm{K}$.}
\end{figure*}

Figure \ref{fig:figure4} shows the evolution of the marginalized probability distributions given in Eq.\ \ref{eq:marginalized_distributions} for a single xenon atom that is initially confined to the left half of a one-dimensional box of length $l = 1.2 \times 10^{-6} \, \mathrm{m}$ at $T = 298 \, \mathrm{K}$. $\rho_{x}(x, t)$ is initially a uniform probability distribution $2 / l$ on the left half of the box and gradually evolves to a uniform distribution $1 / l$ on the entire box. $\rho_{p}(p, t)$ starts as a Maxwell--Boltzmann distribution, but skews to the right shortly after the barrier is removed due to the reflection of particles at $x = 0$. With the further passage of time $\rho_{p}(p, t)$ actually fails to converge as a function at all, but does weakly converge to a Maxwell--Boltzmann distribution as a measure. This is particularly significant since the weak convergence to a Maxwell--Boltzmann distribution occurs without any recourse to the assumption of molecular chaos (the \emph{Stoßzahlansatz}) that Boltzmann believed to be essential in his derivation of the $H$-theorem \cite{maxwell1867iv,boltzmann1872weitere}.

\begin{figure}
	\center
	\includegraphics[width=1.0\linewidth]{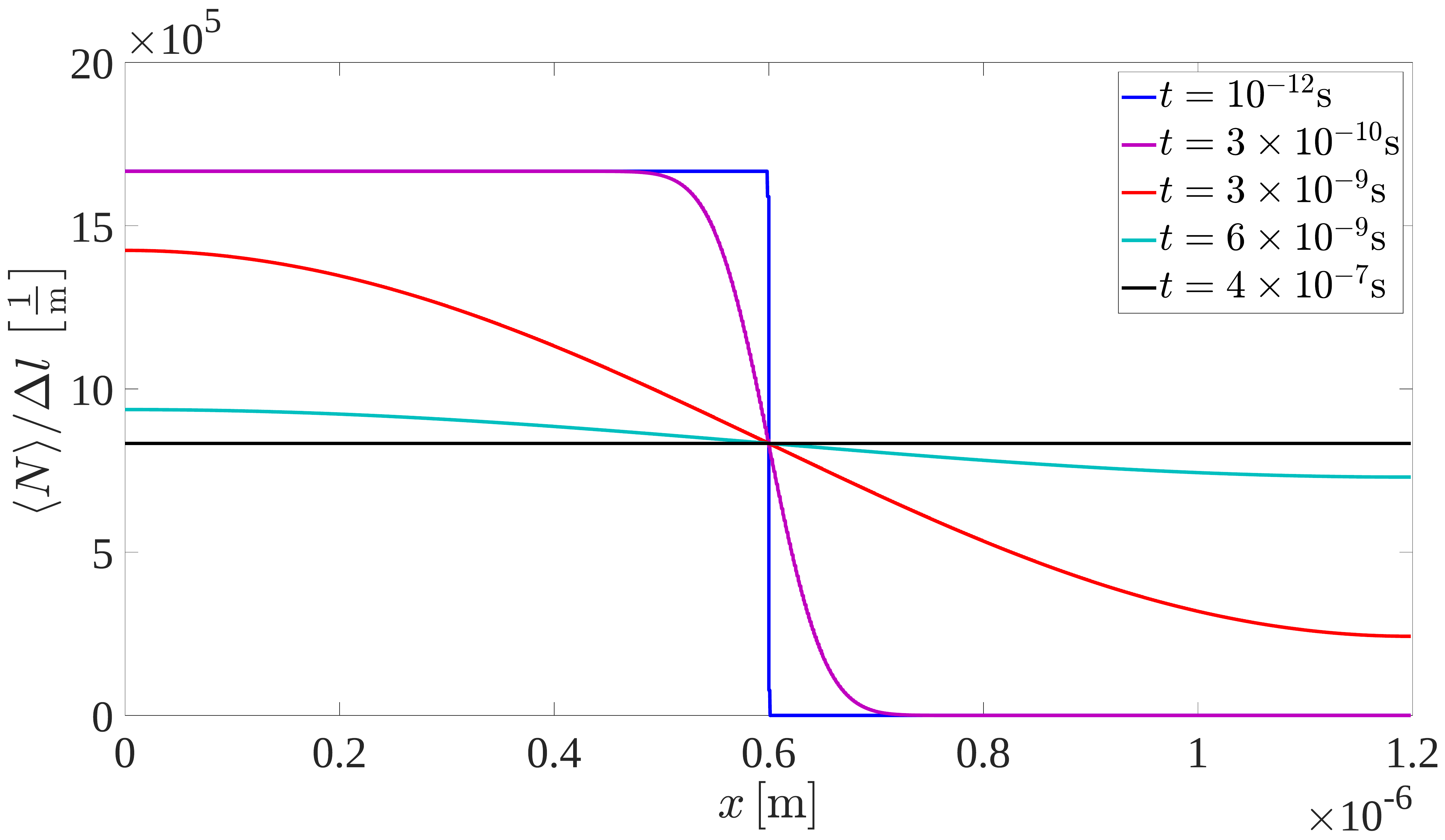}
	\caption{\label{fig:figure5}The evolution of the particle density of the single-particle solution as a function of position. The total number of particles is conserved at all times with a relative error on the order of $10^{-15}$.}
\end{figure}

One of the contributions of Sec.\ \ref{sec:fundamental} are equations to distribute the extensive thermodynamic quantities in Eq.\ \ref{eq:fundamental_classical} over arbitrary spatial partitions. This allows, e.g., densities of extrinsic quantities to be defined without resorting to the standard practice of spatial coarse-graining \cite{balluffi2005kinetics}. Densities are found here by partitioning the one-particle system into $1000$ equal-sized subsystems and plotting the ratio of an extensive quantity and the subsystem width $\Delta l$. For example, the particle density is calculated using Eq.\ \ref{eq:expected_number_per_partition} and plotted as a function of position in Fig.\ \ref{fig:figure5} at various times. The evolution of the particle density closely follows that of $\rho_{x}(x, t)$ in Fig.\ \ref{fig:figure4}, initially being uniformly distributed over the left half of the box and converging to a uniform distribution over the entire box.

\begin{figure}
	\center
	\includegraphics[width=1.0\linewidth]{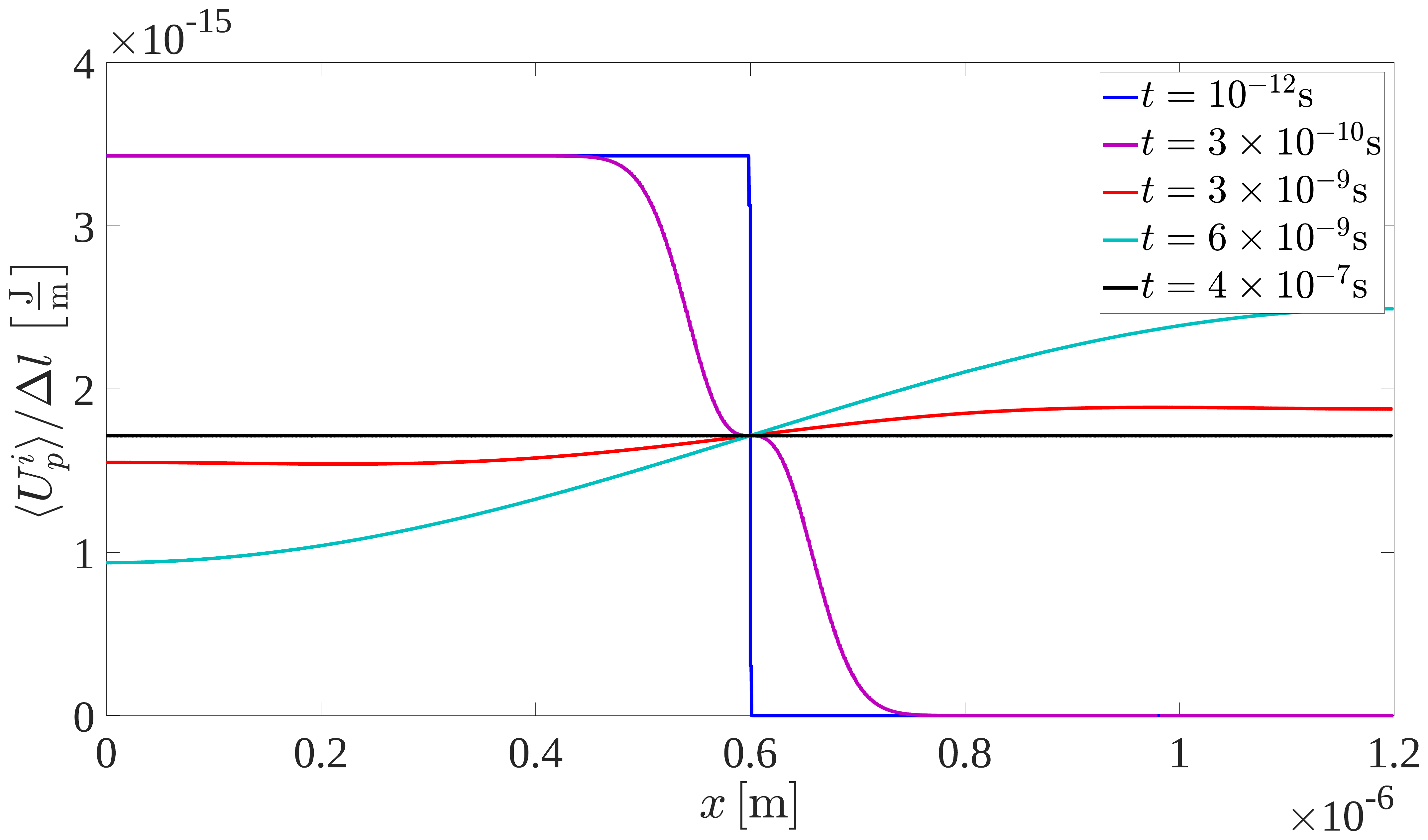}
	\caption{\label{fig:figure6}The evolution of the kinetic energy density of the single-particle solution as a function of position. The total kinetic energy is conserved at all times with a relative error on the order of $10^{-15}$.}
\end{figure}

Since ideal gas particles do not interact with each other and have zero potential energy by definition, only the kinetic energy contribution to the internal energy of a single particle is considered. The kinetic energy density is calculated using Eq.\ \ref{eq:kinetic_energy_part} and plotted as a function of position in Fig. \ref{fig:figure6} at various times. Initially the kinetic energy is uniformly distributed over the left half of the box, though as the expected position of the particle moves to the right the expected kinetic energy of the right half increases. For a brief period the expected kinetic energy on the right actually exceeds that on the left (light blue curve) since a faster (slower) particle has a higher probability of being on the right (left). Predictably, the kinetic energy density eventually converges to a uniform distribution over the entire box.

\begin{figure*}
	\center
	\includegraphics[width=1.0\linewidth]{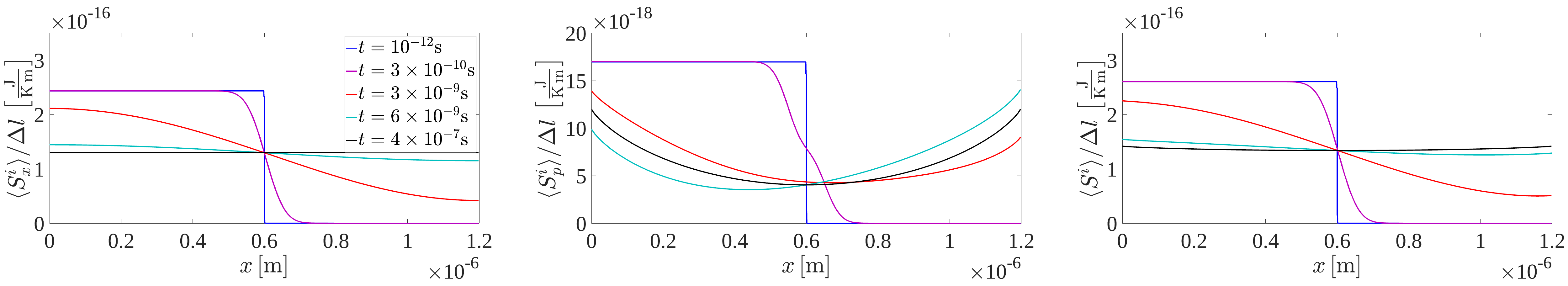}
	\caption{\label{fig:figure7}The evolution of the configurational (left), vibrational (middle), and total (right) entropy densities of the single-particle solution as functions of position.}
\end{figure*}

\begin{figure*}
	\center
	\includegraphics[width=1.0\linewidth]{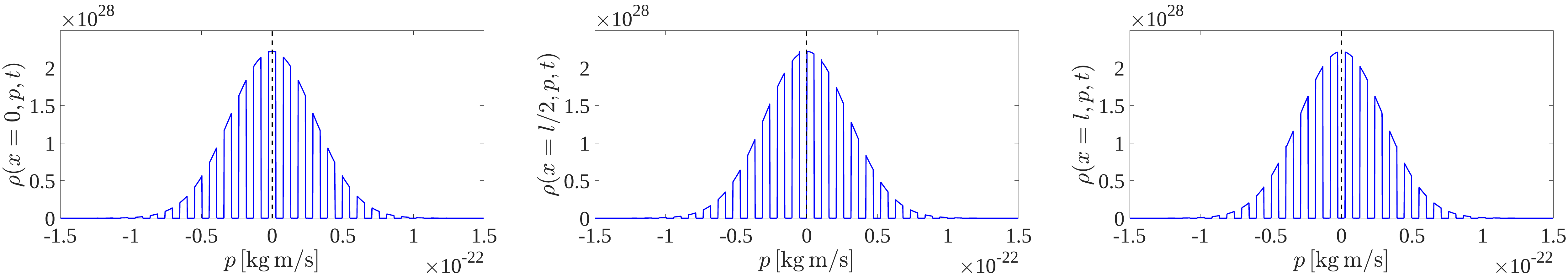}
	\caption{\label{fig:figure8}The probability distribution $\rho(x, p, t)$ of the single-particle solution at $x = 0$ (left), $x = l / 2$ (middle) and $x = l$ (right) at $t = 5 \times 10^{-8} \, \mathrm{s}$.}
\end{figure*}

The densities of the configurational, vibrational, and total entropies are calculated using Eqs.\ \ref{eq:entropy_partition_parts} and \ref{eq:entropy_partition} and are plotted as functions of position in Fig.\ \ref{fig:figure7} at various times. While the evolution of the configurational entropy density closely follows that of the particle density in Fig.\ \ref{fig:figure5}, the behavior of the vibrational entropy is quite different. It is initially distributed uniformly over the left half of the box, but converges to a distribution with higher vibrational entropy density at the ends of the box than in the middle. The precise reason for this is not known, but is apparently related to $\rho(x, p, t)$ not being a constant function of $x$ at any time. For example, $\rho(0, p, t)$ and $\rho(l, p, t)$ are seen in Fig.\ \ref{fig:figure8} to be even functions of $p$ with complementary supports (the sets of possible momenta of a particle at $x = 0$ and $x = l$ are disjoint at all times), whereas the support of $\rho(l / 2, p, t)$ is antisymmetric about the origin. Moreover, while all three of the distributions in Fig.\ \ref{fig:figure8} seem to weakly converge to a Maxwell--Boltzmann distribution, the vibrational entropy density for a Maxwell--Boltzmann distribution would be $8.5 \times 10^{-18} \, \mathrm{J/(K \, m)}$ whereas the observed values are $1.2 \times 10^{-17} \, \mathrm{J/(K \, m)}$ at the ends of the box and $4 \times 10^{-18} \, \mathrm{J/(K \, m)}$ in the middle. This results in the nonuniformity of the converged total entropy density on the right of Fig.\ \ref{fig:figure7}.

\begin{figure*}
	\center
	\includegraphics[width=1.0\linewidth]{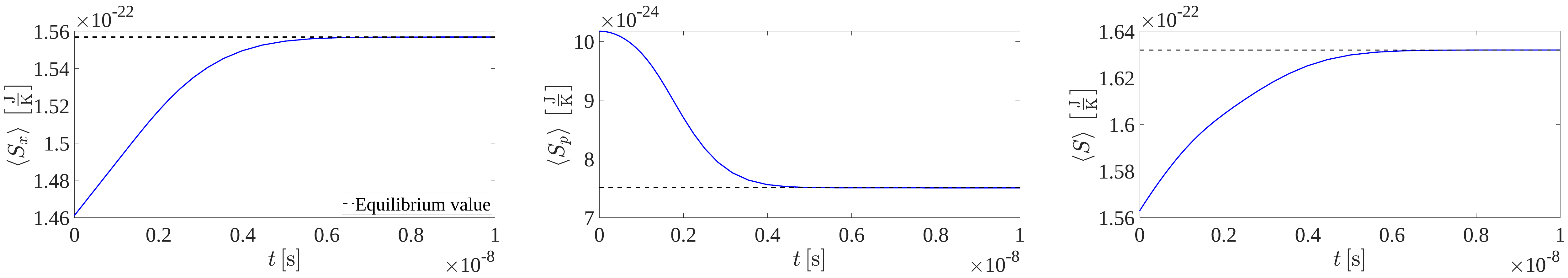}
	\caption{\label{fig:figure9}The configurational (left), vibrational (middle) and the total entropy (right) of the single-particle solution as a function of time.}
\end{figure*}

The integrated configurational, vibrational, and total entropies are shown in Fig.\ \ref{fig:figure9} as functions of time. The Sackur--Tetrode equation predicts that the total entropy should change by $9.5 \times 10^{-24} \, \mathrm{J/K}$ during the gas expansion, precisely equal to the observed increase in the configurational entropy. The Sackur--Tetrode equation does not predict a decrease in the vibrational entropy though, the effect of which is to reduce the total entropy change to $6.9 \times 10^{-24} \, \mathrm{J/K}$; this is a consequence of the Sackur--Tetrode equation assuming that the momentum distribution is unaffected by the expansion of the gas. While the decrease of vibrational entropy is not inconsistent with the second law of thermodynamics (which is conjectured in Sec.\ \ref{subsec:entropy} to apply only to the total entropy), there are several conclusions that could be drawn from this phenomenon.

A first possibility is that the Sackur--Tetrode equation should hold, and that the inconsistency is in the use of Eq.\ \ref{eq:modified_entropy} for the entropy. Indeed, the weak convergence of the probability distribution of momentum to a Maxwell--Boltzmann distribution seems to suggest that the equation for the entropy could be modified to make the vibrational entropy of the two measures the same. Considering Fig.\ \ref{fig:figure8}, this would likely involve a variation on the coarse-graining procedure discussed in Sec.\ \ref{subsec:liouville}. Since the the concerns raised there have not yet been resolved, this line of thought is not considered further.

A second possibility is that the Sackur--Tetrode equation should hold, and the inconsistency be resolved by a variation on Jaynes' forgetting of information. Observe that a practitioner who makes repeated and independent measurements of the position and momentum of the particle would likely conclude that $\rho(x, p, t)$ is a product of a uniform distribution $\rho_{x}(x, t)$ and a Maxwell--Boltzmann distribution $\rho_{p}(p, t)$, and hence that the Sackur--Tetrode equation holds. The argument is effectively that some information about the probability distribution of microstates is lost as a consequence of experimental measurements not being able to resolve the fine structure of the probability distribution of momentum. This seems more reasonable, but raises additional questions about precisely what information should be forgotten and how to do so.

A third possibility (and the one advocated by the authors) is that the Sackur--Tetrode equation does not hold unconditionally, but only when manipulating systems in a classical context. Knowledge of the fine structure of the probability distribution of momentum could, in principle, be used by a microscopic entity resembling Maxwell's demon to extract more work from the system than would be available classically. This would presumably be reflected by a lower entropy than that predicted by the Sackur--Tetrode equation, and could frequently occur when finite systems are concerned. If the practitioner is only able to view the system in a classical context though, then the probability distribution of momentum would likely be mistaken for a Maxwell--Boltzmann distribution as in the previous paragraph and the classical result recovered. Thus the discrepancy with the Sackur--Tetrode equation is not viewed as an inconsistency, but as an indication of the additional opportunities available with a finite non-equilibrium statistical thermodynamics.

\section{Conclusions}
\label{sec:conclusions}

The existing foundations of statistical thermodynamics were developed more than a century ago, at a time before the conception of modern experimental and computational capabilities. Scanning tunneling electron microscopy was not available to resolve the motion of individual atoms, and the calculations required to solve the evolution of nontrivial atomic systems were intractable until the advent of the electronic computer \cite{gibson1960dynamics}. It is entirely understandable that the likes of Maxwell, Boltzmann and Gibbs should be concerned with predicting and explaining the experimental results available at that time, and did not hesitate to use simplifying assumptions like equilibrium and the thermodynamic limit to make the required calculations feasible.

The foundations of statistical thermodynamics have remained essentially unchanged since then though, despite the ongoing efforts of the scientific community. We suspect that the main obstacle to progress is the belief that the thermodynamic ensemble is sacrosanct. The use of thermodynamic ensembles raises questions about the definition of a macrostate and the construction of an ensemble; abandoning the practice instead raises questions about the interpretation of the probability distribution of microstates. Both approaches sow doubts about the reliability of understanding experimental systems by measuring thermodynamic quantities in molecular dynamic simulations, since such simulations are by necessity of finite non-equilibrium systems. Our view is that this situation is no longer tenable.

Since precisely defining thermodynamic ensembles for non-equilibrium states seems to be fraught with difficulties, this paper instead interprets the probability distribution of microstates without using ensembles. This quickly leads to the position advocated by Jaynes \cite{jaynes1992gibbs}, that the probability distribution describes the subjective uncertainty of the practitioner about the microstate. The implication is that thermodynamics is not really concerned with the objective state of a thermodynamic system, but rather with the ability of the practitioner to manipulate that state. The entropy is interpreted in the context of the fundamental thermodynamic relation as indicating the amount of internal energy that cannot be extracted as work as due to the practitioner's imperfect knowledge of the microstate. Consistent resolutions to the mixing paradox, Gibbs' paradox, Loschmidt's paradox, and Maxwell's demon thought experiment are further consequences of the subjectivity of the probability distribution.

An equation for the entropy is proposed in Sec.\ \ref{subsec:entropy} that coincides with Gibbs' entropy for the canonical ensemble, and by the equivalence of ensembles, is consistent with effectively all prior thermodynamic results in the equilibrium thermodynamic limit. The essential difference with Gibbs' entropy is that the proposed definition increases with the mutual information of the position and momentum variables, allowing the dynamics imposed by Liouville's equation to cause the spontaneous increase of entropy in non-equilibrium systems. That is, an answer is offered to the question of irreversibility. Definitions are proposed that allow all of the extensive quantities in the fundamental thermodynamic relation to be distributed over arbitrary thermodynamic subsystems. These definition are used to analyze the expansion of an ideal gas into an isolated box. The results are found to coincide with those expected in a classical context, and suggest that there are additional opportunities offered by a finite non-equilibrium statistical thermodynamics.

The finite non-equilibrium statistical thermodynamics begun here is not complete. Notably, the intrinsic quantities appearing in the fundamental thermodynamic relation are not addressed, and there is reason to believe that their definition will follow along different lines than those explored here. Their development is intended for a future publication.

\begin{acknowledgments}
	OBE and JKM were supported by the National Science Foundation under Grant No.\ DMR 1839370.
\end{acknowledgments}

\appendix

\section{Partitioning Potential Energy}
\label{sec:partitioning_potential_energy}

\begin{figure}
	\center
	\includegraphics[width=0.8\linewidth]{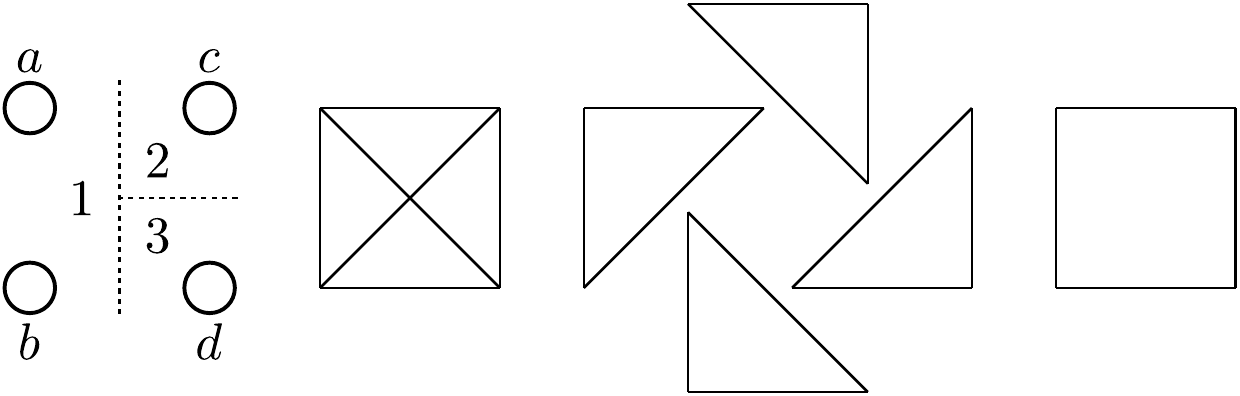}
	\caption{\label{fig:figure10}A partition of a four-atom system into three subsystems with the subsystem and atom labels on the left. The potential energy includes contributions from four $1$-body, six $2$-body, four $3$-body, and one $4$-body interactions, represented graphically on the right.}
\end{figure}

Consider the example system in Fig.\ \ref{fig:figure10} where there are four atoms and three partitions. Table \ref{tbl:weights} shows all of the multiplicative factors $\abs{\mathcal{L}^{i} \cap s} / \abs{s}$ that appear in Eq.\ \ref{eq:potential_energy_part}. Similarly, Table \ref{tbl:coefficients} shows the coefficients $a_{rs}$ for events that remove atoms with labels in $r$. The coefficients form a $15 \times 15$ invertible matrix for four atoms, allowing potential energies to be assigned to the three partitions by means of Eq.\ \ref{eq:potential_energy_part2}.

\begin{table*}
	\center
	\begin{tabular}{c c c c c c c c c c c c c c c c}
		$ i $ & $\phi_{a}$ & $\phi_{b}$ & $\phi_{c}$ & $\phi_{d}$ & $\phi_{ab}$ & $\phi_{ac}$ & $\phi_{ad}$ & $\phi_{bc}$ & $\phi_{bd}$ & $\phi_{cd}$ & $\phi_{abc}$ & $\phi_{abd}$ & $\phi_{acd}$ & $\phi_{bcd}$ & $\phi_{abcd}$\\
		\hline\hline
		$1$ & $1$ & $1$ & $0$ & $0$ & $1$ & $1/2$ & $1/2$ & $1/2$ & $1/2$ & $0$ & $2/3$ & $2/3$ & $1/3$ & $1/3$ & $1/2$ \\
		$2$ & $0$ & $0$ & $1$ & $0$ & $0$ & $1/2$ & $0$ & $1/2$ & $0$ & $1/2$ & $1/3$ & $0$ & $1/3$ & $1/3$ & $1/4$ \\
		$3$ & $0$ & $0$ & $0$ & $1$ & $0$ & $0$ & $1/2$ & $0$ & $1/2$ & $1/2$ & $0$ & $1/3$ & $1/3$ & $1/3$ & $1/4$ \\ 
	\end{tabular}
	\caption{The weights $\abs{\mathcal{L}^{i} \cap s} / \abs{s}$ associated with each $\phi_{s}$ for the example system in Fig.\ \ref{fig:figure10}.}
	\label{tbl:weights}
\end{table*}

\begin{table*}
	\center
	\begin{tabular}{c c c c c c c c c c c c c c c c}
		$ $ & $\phi_{a}$ & $\phi_{b}$ & $\phi_{c}$ & $\phi_{d}$ & $\phi_{ab}$ & $\phi_{ac}$ & $\phi_{ad}$ & $\phi_{bc}$ & $\phi_{bd}$ & $\phi_{cd}$ & $\phi_{abc}$ & $\phi_{abd}$ & $\phi_{acd}$ & $\phi_{bcd}$ & $\phi_{abcd}$\\
		\hline\hline
		$\chi_{a}$ & $-1$ & $0$ & $0$ & $0$ & $-1$ & $-1$ & $-1$ & $0$ & $0$ & $0$ & $-1$ & $-1$ & $-1$ & $0$ & $-1$\\
		$\chi_{b}$ & $0$ & $-1$ & $0$ & $0$ & $-1$ & $0$ & $0$ & $-1$ & $-1$ & $0$ & $-1$ & $-1$ & $0$ & $-1$ & $-1$\\
		$\chi_{c}$ & $0$ & $0$ & $-1$ & $0$ & $0$ & $-1$ & $0$ & $-1$ & $0$ & $-1$ & $-1$ & $0$ & $-1$ & $-1$ & $-1$\\
		$\chi_{d}$ & $0$ & $0$ & $0$ & $-1$ & $0$ & $0$ & $-1$ & $0$ & $-1$ & $-1$ & $0$ & $-1$ & $-1$ & $-1$ & $-1$\\
		$\chi_{ab}$ & $-1$ & $-1$ & $0$ & $0$ & $-1$ & $-1$ & $-1$ & $-1$ & $-1$ & $0$ & $-1$ & $-1$ & $-1$ & $-1$ & $-1$\\
		$\chi_{ac}$ & $-1$ & $0$ & $-1$ & $0$ & $-1$ & $-1$ & $-1$ & $-1$ & $0$ & $-1$ & $-1$ & $-1$ & $-1$ & $-1$ & $-1$\\
		$\chi_{ad}$ & $-1$ & $0$ & $0$ & $-1$ & $-1$ & $-1$ & $-1$ & $0$ & $-1$ & $-1$ & $-1$ & $-1$ & $-1$ & $-1$ & $-1$\\
		$\chi_{bc}$ & $0$ & $-1$ & $-1$ & $0$ & $-1$ & $-1$ & $0$ & $-1$ & $-1$ & $-1$ & $-1$ & $-1$ & $-1$ & $-1$ & $-1$\\
		$\chi_{bd}$ & $0$ & $-1$ & $0$ & $-1$ & $-1$ & $0$ & $-1$ & $-1$ & $-1$ & $-1$ & $-1$ & $-1$ & $-1$ & $-1$ & $-1$\\
		$\chi_{cd}$ & $0$ & $0$ & $-1$ & $-1$ & $0$ & $-1$ & $-1$ & $-1$ & $-1$ & $-1$ & $-1$ & $-1$ & $-1$ & $-1$ & $-1$\\
		$\chi_{abc}$ & $-1$ & $-1$ & $-1$ & $0$ & $-1$ & $-1$ & $-1$ & $-1$ & $-1$ & $-1$ & $-1$ & $-1$ & $-1$ & $-1$ & $-1$\\
		$\chi_{abd}$ & $-1$ & $-1$ & $0$ & $-1$ & $-1$ & $-1$ & $-1$ & $-1$ & $-1$ & $-1$ & $-1$ & $-1$ & $-1$ & $-1$ & $-1$\\
		$\chi_{acd}$ & $-1$ & $0$ & $-1$ & $-1$ & $-1$ & $-1$ & $-1$ & $-1$ & $-1$ & $-1$ & $-1$ & $-1$ & $-1$ & $-1$ & $-1$\\
		$\chi_{bcd}$ & $0$ & $-1$ & $-1$ & $-1$ & $-1$ & $-1$ & $-1$ & $-1$ & $-1$ & $-1$ & $-1$ & $-1$ & $-1$ & $-1$ & $-1$\\
		$\chi_{abcd}$ & $-1$ & $-1$ & $-1$ & $-1$ & $-1$ & $-1$ & $-1$ & $-1$ & $-1$ & $-1$ & $-1$ & $-1$ & $-1$ & $-1$ & $-1$\\
	\end{tabular}
	\caption{The coefficients $a_{rs}$ for the example system in Fig.\ \ref{fig:figure10}.}
	\label{tbl:coefficients}
\end{table*}

\section{Formula for Weakly Extrinsic Potential Energy}
\label{sec:extrinsic_formula}

The proof that Eq.\ \ref{eq:potential_energy_part3} distributes the potential energy of a system $A$ in the desired way over the $n$ subsystems $A^{i}$ begins with the assumption that the potential energy can be written as a sum of many-body interactions. Consider a single $m$-body interaction with atoms contained in $p$ subsystems. The hypothesis is that the number of terms on the right hand side of Eq.\ \ref{eq:potential_energy_part3} containing this interaction is precisely $n! / p$, or that the first time an atom participating in the interaction is removed coincides with removing all the atoms of $A^{i}$ for $n! / p$ of the $n!$ possible permutations. This would result in the overall contribution of the interaction to $U^{i}_{x}(\vec{x})$ being $1 / p$ times the energy of the interaction, as desired.

Consider the number of permutations such that the $q$ subsystems selected before $A^{i}$ do not participate in the interaction. There are $(n - p)! / (n - p - q)!$ ways to select such initial subsystems, and $(n - 1 - q)!$ ways to select the subsystems afterwards. This implies that the hypothesis is true if the equation
\begin{equation*}
\frac{n!}{p} = \sum_{q = 0}^{n - p} \frac{(n - p)!}{(n - p - q)!} (n - 1 - q)!
\end{equation*}
holds. This can be shown to be so by making the substitution $r = n - p - q$ and replacing the summand with a binomial coefficient to find
\begin{equation*}
\frac{n!}{p} = (n - p)! (p - 1)! \sum_{r = 0}^{n - p} \binom{p - 1 + r}{r}.
\end{equation*}
The summation can be replaced by a single binomial coefficient using a variant of the Chu--Vandermonde identity with the result
\begin{equation*}
\frac{n!}{p} = (n - p)! (p - 1)! \binom{n}{n - p}.
\end{equation*}
Expanding the binomial coefficient and canceling the resulting terms reduces this to an identity.

\section{Evolution of an Ideal Gas in a Box}
\label{sec:ideal_gas_in_a_box}

Liouville's equation in principle describes the evolution of the probability distribution $\rho(\vec{x}, \vec{p}, t)$ for any isolated system, but is difficult to solve in practice. An ideal gas in an isolated box is one of the few systems for which a general solution is known \cite{hobson1968exact}, and relies on the observation that the problem can be reduced to a single particle moving in a one-dimensional box. The general solution can be written as a product of the one-dimensional solutions, or
\begin{align}
	\rho(\vec{x},\vec{p},t) &= \prod_{j=1}^{N} \prod_{\mu=1}^{d} \rho(x_{\mu j},p_{\mu j},t) \nonumber \\
	\rho(x_{\mu j},p_{\mu j},t) &= \text{e}^{-i \hat{L}_{\mu j} t} \rho(x_{\mu j},p_{\mu j},0)
	\label{eq:general_solution_as_product}
\end{align}
where $j$ and $\mu$ index the particles and spatial dimensions and $\hat{L}_{\mu j}$ is the one-dimensional Liouville operator. A procedure to obtain the one-dimensional solution was developed by Born \cite{born1955continuity} using the method of images. This replaces the actual Hamiltonian $H$ by the free-particle Hamiltonian $\bar{H} = p^2 / {2 m}$ (the subscripts $\mu$ and $j$ are dropped for simplicity) and handles the boundary conditions by modifying the probability distribution. The initial probability distribution $\rho(x,p,0)$ is replaced by an extended probability distribution $\bar{\rho}(x, p, 0)$ such that $\bar{\rho}(-x, -p, 0) = \bar{\rho}(x, p, 0)$ and $\bar{\rho}(x + 2 l, p, 0) = \bar{\rho}(x, p, 0)$ where $l$ is the length of the box. The extended problem is easier to solve, and the solution of the actual problem is $\rho(x, p, t) = \bar{\rho}(x, p, t)$ for $0 < x < l$. Details of this approach can be found in Refs.\ \cite{hobson1968exact,born1955continuity}.

Consider a system of $N$ ideal gas particles initially confined to the left-hand side of a one-dimensional isolated box. The gas is in thermodynamic equilibrium with the initial condition
\begin{align*}
	\rho(x,p,0) =
	\begin{cases}
		\rho_{x}(x,0) \rho_{p}(p,0)& \text{for $0 < x < l'$},\\
		0& \text{for $l' < x < l$}.
	\end{cases}
\end{align*}
where $\rho_{x}(x,0) = 1 / l'$ is a uniform probability distribution and $\rho_{p}(p,0)$ is the Maxwell--Boltzmann distribution. The extended probability distribution $\bar{\rho}(x,p,0)$ is constructed as above, and along with the free-particle Hamiltonian $\bar{H}$ gives \cite{hobson1968exact}
\begin{equation}
	\bar{\rho}(x,p,t) = \bar{\rho}(x - p t,p,0)
	\label{eq:ideal_gas_solution}
\end{equation}
for the time evolution. Since $\bar{\rho}(x,p,0)$ is an even function of $x$ it can be written as the Fourier cosine series
\begin{align*}
\bar{\rho}(x,p,0) &= a_{0}(p) + \sum_{n=1}^{\infty} a_{n}(p) \cos{\Big( \frac{n \pi x } {l}\Big)} \\
	a_{0}(p) &= \frac{1}{l} \rho_{p}(p,0) \\
	a_{n}(p) &= \frac{2}{n \pi l'} \sin{\Big( \frac{n \pi l' } {l}\Big)} \rho_{p}(p,0).
\end{align*}
Using Eq.\ \ref{eq:ideal_gas_solution} then yields the solution of a single particle with the free-particle Hamiltonian as
\begin{align*}
	\bar{\rho}(x, p, t) = \Bigg\{ \frac{1}{l} + \frac{2}{\pi l'} \sum_{n=1}^{\infty} & \frac{1}{n} \sin{\Big( \frac{n \pi l'}{l} \Big)} \\
	& \times \cos{ \Big[ \frac{n \pi}{l} \Big(x - \frac{p t}{m}\Big) \Big]} \Bigg\} \rho_{p}(p, 0) .
\end{align*}
Recall that the solution of the actual problem is
\begin{equation}
	\rho(x,p,t) = \bar{\rho}(x,p,t) \quad \text{for} \quad 0 < x < l.
	\label{eq:single_ideal_gas_solution}
\end{equation}
Along with Eq.\ \ref{eq:general_solution_as_product}, this can be used to find, e.g., the solution for $N$ particles initially confined to the left-hand side of a three-dimensional box. The single particle marginal probability distributions $\rho_{x}(x,t)$ and $\rho_{p}(p,t)$ are
\begin{align}
	\rho_{x}(x,t) &= \bigg[ \frac{1}{l} + \frac{2}{\pi l'} \sum_{n=1}^{\infty} \frac{1}{n} \sin{\Big( \frac{n \pi l'}{l} \Big)} \exp{\Big( -\frac{n^2 t^2}{2 \sigma^2} \Big)} \nonumber \\
	&\mkern 130mu \times \cos{\Big( \frac{n \pi x}{l} \Big)} \bigg], \nonumber \\
	\rho_{p}(p,t) &= \bigg[ 1 + \frac{4 l}{\pi^2 l'} \sum_{n=1}^{\infty}{\vphantom{\sum}}' \frac{1}{n^2} \sin{\Big( \frac{n \pi l'}{l} \Big)} \nonumber \\
	&\mkern 130mu \times \sin{\Big( \frac{n \pi p t}{l m} \Big)} \bigg] \rho_{p}(p,0)
	\label{eq:marginalized_distributions}
\end{align}
where $\sigma = l / \pi \sqrt{m / (k_B T)}$ and the primed summation is only over odd $n$. With increasing time the exponential term in $\rho_{x}(x,t)$ goes to zero and $\rho_{x}(x,t)$ converges to $1 / l$ as expected. The dependence of $\rho_{p}(p,t)$ on time suggests that the limit is not well-defined as a function, but $\rho_{p}(p,t)$ does appear to weakly converge to a Maxwell--Boltzmann distribution as a measure \cite{hobson1968exact}.


\vanish{
[12] Fuyang Tian, Lorand Delczeg, Nanxian Chen, Lajos Karoly Varga, Jiang Shen, and
Levente Vitos. Structural stability of NiCoFeCrAl x high-entropy alloy from ab initio
theory. Phys. Rev. B, 88:085128, Aug 2013.
[13] Chuan Zhang, Fan Zhang, Shuanglin Chen, and Weisheng Cao. Computational ther-
modynamics aided high-entropy alloy design. JOM, 64(7):839–845, 2012.
[15] Pasquale Pavone, Stefano Baroni, and Stefano de Gironcoli. ab phase transition in
tin: A theoretical study based on density-functional perturbation theory. Phys. Rev.
B, 57:10421–10423, May 1998.
[16] Jürgen Neuhaus, Michael Leitner, Karl Nicolaus, Winfried Petry, Bernard Hennion,
and Arno Hiess. Role of vibrational entropy in the stabilization of the high-temperature
phases of iron. Phys. Rev. B, 89:184302, May 2014.
[26] A. van de Walle and G. Ceder. First-principles computation of the vibrational entropy
of ordered and disordered Pd 3 V. Phys. Rev. B, 61:5972–5978, Mar 2000.
[38] W. Jones and N.H. March. Theoretical Solid State Physics: Perfect lattices in equilibrium. Dover Books on Physics Series. Dover Publications, 1973.
[39] H. Peng, C.L. Wang, J.C. Li, H.C. Wang, Y. Sun, and Q. Zheng. Elastic and vibra-
tional properties of Mg 2 Si 1-x Sn x alloy from first principles calculations. Solid State Communications, 152(9):821–824, 2012.
[40] Olle Eriksson, J. M. Wills, and Duane Wallace. Electronic, quasiharmonic, and anhar-
monic entropies of transition metals. Phys. Rev. B, 46:5221–5228, Sep 1992.
[41] P. Souvatzis and S. P. Rudin. Dynamical stabilization of cubic ZrO 2 by phonon-phonon
interactions: Ab initio calculations. Phys. Rev. B, 78:184304, Nov 2008.
[108] Nicholas Metropolis, Arianna W. Rosenbluth, Marshall N. Rosenbluth, Augusta H.
Teller, and Edward Teller. Equation of state calculations by fast computing machines.
The Journal of Chemical Physics, 21(6), 1953.
[109] Fugao Wang and D. P. Landau. Efficient, multiple-range random walk algorithm to
calculate the density of states. Phys. Rev. Lett., 86:2050–2053, Mar 2001.
[110] Daan Frenkel and Anthony J. C. Ladd. New Monte Carlo method to compute the free
energy of arbitrary solids. application to the fcc and hcp phases of hard spheres. The
Journal of Chemical Physics, 81(7):3188–3193, 1984.
[115] M. de Koning and A. Antonelli. Einstein crystal as a reference system in free energy
estimation using adiabatic switching. Phys. Rev. E, 53:465–474, Jan 1996.
}
	
\end{document}